\documentclass[twocolumn]{aastex702}

\usepackage{natbib}
\usepackage{amsmath}
\usepackage{multirow}
\usepackage[normalem]{ulem}
\usepackage{xcolor}
\usepackage{xspace}
\usepackage{enumitem}   
\usepackage{graphicx}
\usepackage{wrapfig}
\usepackage[super]{nth}
\usepackage{hyperref}
\usepackage{url}

\newcommand{\Rearth}{$R_\oplus$\xspace}
\newcommand{\Mearth}{$M_\oplus$\xspace}

\newcommand{\Msun}{$M_\odot$\xspace}

\newcommand{\tess}{\texttt{TESS}\xspace}

\newcommand\vsini{$v$~sin~$i$}

\newcommand{\unit}[1]{\ensuremath{\, \mathrm{#1}}}
\newcommand{\arcsecond}{$''$\xspace}
\definecolor{tropicalrainforest}{rgb}{0.0, 0.46, 0.37}
\definecolor{plum}{HTML}{88498f}

\shorttitle{TOI-2120}
\shortauthors{Fernandes et al.}
\newcommand{\PSUAA}{Department of Astronomy and Astrophysics, 525 Davey Laboratory, 251 Pollock Road, Penn State, University Park, PA, 16802, USA}
\newcommand{\PSUCEHW}{Center for Exoplanets and Habitable Worlds, 525 Davey Laboratory, 251 Pollock Road, Penn State, University Park, PA, 16802, USA}

\newcommand{\UA}{Steward Observatory, University of Arizona, 933 N.\ Cherry Ave, Tucson, AZ 85721, USA}

\newcommand{\Caltech}{Department of Astronomy, California Institute of Technology, 1200 E California Blvd, Pasadena, CA 91125, USA}

\newcommand{\GoddardESAL}{Exoplanets and Stellar Astrophysics Laboratory, NASA Goddard Space Flight Center, Greenbelt, MD 20771, USA}

\newcommand{\Macquarie}{School of Mathematical and Physical Sciences, Macquarie University, Balaclava Road, North Ryde, NSW 2109, Australia}

\newcommand{\CUBoulderEE}{Electrical, Computer \& Energy Engineering, 440 UCB, University of Colorado, Boulder, CO 80309, USA}
\newcommand{\CUBoulder}{Department of Physics, 390 UCB, University of Colorado, Boulder, CO 80309, USA}
\newcommand{\JPL}{Jet Propulsion Laboratory, California Institute of Technology, 4800 Oak Grove Drive, Pasadena, California 91109}

\newcommand{\UCI}{Department of Physics \& Astronomy, The University of California, Irvine, Irvine, CA 92697, USA}

\newcommand{\Carnegie}{Earth and Planets Laboratory, Carnegie Science, 5241 Broad Branch Road, NW, Washington, DC 20015, USA}
\newcommand{\PSUICS}{Institute for Computational and Data Sciences, Penn State, University Park, PA, 16802, USA}

\newcommand{\TIFR}{Department of Astronomy and Astrophysics, Tata Institute of Fundamental Research, Homi Bhabha Road, Colaba, Mumbai 400005, India}

\begin{document}
\title{Revisiting TOI-2120 b: Updated mass and radius constraints on a sub-Neptune orbiting an M4 dwarf\footnote{Based on observations obtained with the Hobby-Eberly Telescope (HET), which is a joint project of the University of Texas at Austin, the Pennsylvania State University, Ludwig-Maximillians-Universitaet Muenchen, and Georg-August Universitaet Goettingen. The HET is named in honor of its principal benefactors, William P. Hobby and Robert E. Eberly}}

% \correspondingauthor{Rachel B. Fernandes}
% \email{rbf5378@psu.edu}

\author[0000-0002-3853-7327]{Rachel B. Fernandes}
\altaffiliation{Center for Exoplanets and Habitable Worlds (CEHW) Fellow}
\affiliation{\PSUAA}
\affiliation{\PSUCEHW}
\email[show]{rbf5378@psu.edu}

\author[0000-0002-9082-6337]{Andrea S.J.\ Lin}
\affiliation{\Caltech}
\email{asjlin@caltech.edu}

\author[0000-0003-0149-9678]{Paul Robertson}
\altaffiliation{Instrument Team Project Scientist}
\affiliation{\UCI}
\email{paul.robertson@uci.edu}

\author[0000-0001-8401-4300]{Shubham Kanodia}
\affiliation{\Carnegie}
\email{skanodia@carnegiescience.edu}

\author[0000-0002-1092-2995]{Anne Dattilo}
\affiliation{\PSUAA}
\affiliation{\PSUCEHW}
\affiliation{\PSUICS}
\email{axd5702@psu.edu}

\author[0000-0003-1439-2781]{Megan Delamer}
\affiliation{\PSUAA}
\affiliation{\PSUCEHW}
\email{mmd6393@psu.edu}

\author[0000-0001-9307-8170]{Brock Parker}
\affiliation{Steward Observatory, The University of Arizona, 933 N. Cherry Avenue, Tucson, AZ 85721, USA}
\email{baparker@arizona.edu}

\author[0000-0003-4835-0619]{Caleb I. Ca\~nas}
\affiliation{Southeastern Universities Research Association, Washington, DC 20005, USA}
\affiliation{\GoddardESAL}
\email{c.canas@nasa.gov}

\author[0000-0002-6629-4182]{Suzanne Hawley}
\affiliation{Department of Astronomy, Box 351580, University of Washington, Seattle, WA 98195, USA}
\email{slhawley@uw.edu}

% \author[0000-0003-1263-8637]{Leslie Hebb}
% \affiliation{Department of Physics, Hobart and William Smith Colleges, 300 Pulteney Street, Geneva, NY 14456, USA}
% \email{hebb@hws.edu}

\author[0000-0002-4475-4176]{Henry A. Kobulnicky}
\affiliation{Department of Physics \& Astronomy, University of Wyoming, Laramie, WY 82070, USA}
\email{ChipK@uwyo.edu}

% \author[0000-0001-7458-1176]{Adam F. Kowalski}
% \affiliation{Department of Astrophysical and Planetary Sciences, University of Colorado Boulder, 2000 Colorado Avenue, Boulder, CO 80305, USA}
% \affiliation{25 National Solar Observatory, University of Colorado Boulder, 3665 Discovery Drive, Boulder, CO 80303, USA}
% \affiliation{26 Laboratory for Atmospheric and Space Physics, University of Colorado Boulder, 3665 Discovery Drive, Boulder, CO 80303, USA}
% \email{adam.f.kowalski@colorado.edu}

\author[0000-0001-9662-3496]{William D. Cochran}
\affiliation{McDonald Observatory and Center for Planetary Systems Habitability, The University of Texas at Austin, Austin, TX 78730, USA}
\email{wdc@astro.as.utexas.edu}

\author[0000-0002-2144-0764]{Scott A.\ Diddams}
\affiliation{\CUBoulderEE}
\affiliation{\CUBoulder}
\email{scott.diddams@colorado.edu}

\author[0000-0003-1312-9391]{Samuel Halverson}
\affiliation{\JPL}
\email{samuel.halverson@jpl.nasa.gov}

\author[0000-0001-8720-5612]{Joe P.\ Ninan}
\affiliation{\TIFR}
\email{indiajoe@gmail.com}

\author[0000-0003-1324-0495]{Leonardo A.\ Paredes}
\affiliation{\UA}
\email{lparedes@arizona.edu}

\author[0000-0002-4046-987X]{Christian Schwab}
\affiliation{\Macquarie}
\email{mail.chris.schwab@gmail.com}

\begin{abstract}
We present updated mass and radius measurements of TOI-2120 b, a short-period (5.80\,days) planet orbiting a mid-M dwarf identified by the Transiting Exoplanet Survey Satellite (TESS). TOI-2120 b was previously validated through transit photometry and high-resolution imaging, with a radius near the upper boundary of the radius valley around low-mass stars. We refined the planetary properties by combining archival TESS and ground-based photometry with new transit observations from the ARC 3.5m and Red Buttes Observatory (RBO) telescopes and new precision RV measurements obtained with the Habitable-zone Planet Finder (HPF) spectrograph. Our joint analysis yields a radius of $2.31\pm0.7$\,\Rearth, larger than previously reported. We also measure a planetary mass of $5.81^{+2.94}_{-2.69}$\,\Mearth at 2$\sigma$, corresponding to a bulk density of $2.62^{+1.37}_{-1.22}$ g cm$^{-3}$, positioning TOI-2120\,b toward the water-rich region predicted by isocomposition models. Additional RV observations will be needed to further constrain its mass and composition.
% TOI-2120 b is among a small number of sub-Neptunes orbiting very cool stars with both mass and radius measurements.
\end{abstract}

\section{Introduction} \label{sec:intro}
Demographic studies from the Kepler \citep{Borucki2010} and K2 \citep{Howell2011} missions revealed that short-period ($<$100\,days) planets between the sizes of Earth and Neptune are common around Sun-like (FGK) stars and that their radii do not follow a continuous distribution. Instead, the population exhibits a bimodal radius distribution, commonly referred to as the radius valley, with a deficit of planets at $\sim$1.5--2\,\Rearth separating smaller super-Earths from larger sub-Neptunes \citep{fulton2017california,hardegree2020scaling}. Subsequent studies showed that the location of the radius valley varies with orbital period and stellar mass \citep{berger2020, Petigura2022}.

Because of their survey design, Kepler and K2 observed a combined sample of only $\sim33,000$ M dwarfs and consequently detected relatively few planets around those stars. The Transiting Exoplanet Survey Satellite (TESS; \citealt{ricker2015_tess}), in contrast, surveys nearly the entire sky and has identified a rapidly growing population of planets around M dwarfs, enabling the radius distribution to be examined as a function of stellar mass. Early studies found evidence for a radius valley around early M dwarfs, but its behavior does not appear to follow the same scaling observed around FGK stars. For example, \citet{cloutiermenou2020} found that the radius valley around cooler stars ($<$4700\,K) shifts toward larger planetary radii with increasing incident flux, opposite to the trend measured for FGK stars. They argued that this behavior could favor formation scenarios that produce planets with different bulk compositions rather than models in which atmospheric mass loss is the dominant mechanism shaping the radius distribution. More recently, \citet{gillis2026} analyzed 8134 mid- to late-M dwarfs observed by TESS and identified 77 transiting planet candidates. They found a unimodal planetary radius distribution ($<$6\,\Rearth) rather than the bimodal distribution observed around higher mass stars, suggesting that the radius valley disappears for lower mass stars.

The origin of the radius bimodality remains under debate. Atmospheric mass-loss models begin with a population of rocky cores that accrete H/He envelopes and subsequently lose these atmospheres through photoevaporation \citep{owenwu2017} and/or core-powered mass loss \citep{ginzburg2018core, gupta2019sculpting}. Formation models instead predict that planets can emerge with substantially different bulk compositions. Global formation and migration models show that planets forming beyond the water ice line can accrete substantial amounts of icy material before migrating inward, producing water-rich planets at short orbital periods \citep[e.g.,][]{alibert2013, raymond2018water, Bitsch2016, Venturini2020} that can occupy the same radius range as sub-Neptunes with H/He envelopes. The distinction between these scenarios is particularly relevant for M dwarfs because their lower luminosities place the ice line closer to the star, altering the composition of material available during planet formation and the subsequent migration pathways of close-in planets. At the same time, the strong irradiation received by short-period planets around M dwarfs can drive substantial atmospheric evolution, potentially modifying the observable radius distribution after formation  \citep[e.g.,][]{Peacock2020,RicheyYowell2023}.

In short, several of these scenarios can produce similar planetary radii despite fundamentally different bulk compositions. Mass measurements therefore provide a critical second dimension: combined with radius, they yield the bulk density and can distinguish between populations with similar radii but different compositions. \citet{luquepalle2022} demonstrated the importance of this distinction by compiling precise mass and radius measurements for 43 small transiting planets around early M dwarfs. They identified three broad populations in mass-radius space: high-density planets consistent with predominantly rocky compositions, lower-density planets consistent with substantial water-rich material, and still lower-density sub-Neptunes requiring significant H/He envelopes. Formation models such as \citet{venturini2024} provide a possible explanation for this structure, showing that pebble accretion and subsequent atmospheric evolution naturally produce an increasing fraction of water-rich planets toward lower stellar masses. In these models, the water-rich population becomes sufficiently abundant to fill the radius valley while retaining a distinct bulk-density distribution, causing the radius valley to weaken or disappear even though the underlying compositional populations remain distinct. This provides a possible explanation for the demographic transition reported by \citet{gillis2026}, where the increasing prevalence of water-rich planets toward lower stellar masses obscures the radius valley. Testing this interpretation, however, requires individual planets with sufficiently precise masses and radii to determine whether they are predominantly rocky, water-rich, or dominated by H/He envelopes.

\citet{Hori2024} identified TOI-2120\,b as a transiting sub-Neptune orbiting a $0.211\pm0.005$\,\Msun M dwarf and determined a planetary radius of $2.122\pm0.065$\,\Rearth and an orbital period of $5.7998164\pm0.0000035$\,days. Their radial velocity (RV) observations yielded an upper mass limit of $<$6.8\,\Mearth, leaving its bulk density and composition poorly constrained \citep{Hori2024}. With a radius in the sub-Neptune regime and a host star near the low-mass end of the M dwarf population, TOI-2120\,b lies in a region where water-rich and H/He-rich compositions have substantially different implications for its formation and subsequent evolution.

In this work, we present updated mass and radius measurements of the previously identified sub-Neptune TOI-2120\,b using new ground-based transit photometry from the 0.6\,m telescope at Red Buttes Observatory (RBO) and the ARC 3.5\,m telescope at Apache Point Observatory (APO), together with new high-resolution spectra and RV measurements from the Habitable-zone Planet Finder (HPF). In Section~\ref{sec:obs}, we describe the photometric and spectroscopic observations used in our analysis. In Section~\ref{sec:activity}, we characterize the host star and investigate stellar activity using the HPF spectra, including its potential contribution to the observed RV variations. In Section~\ref{sec:joint}, we present the joint analysis of the transit and RV data and derive updated planetary and orbital parameters. In Section~\ref{sec:discussion}, we compare our measurements with those of \citet{Hori2024} and place TOI-2120\,b in the context of the emerging population of small planets around M dwarfs. Finally, in Section~\ref{sec:summary}, we summarize our results and their implications for the characterization of TOI-2120\,b.

\section{Observations} \label{sec:obs}
\subsection{TESS Photometry}

TESS observed TOI-2120 (TIC 389900760) in Sectors 18, 24, 25, 52, and 58 (2019 November 3–27, 2020 April 16–May 12, 2020 May 14–June 8, 2022 May 18–June 13, and 2022 October 29–November 26, respectively), with all observations obtained at 2-minute cadence. Figure~\ref{fig:transits} shows the TESS transit signal phase-folded using the combined observations from all sectors. We use the Pre-search Data Conditioning Simple Aperture Photometry (PDCSAP; \citealt{Jenkins2016}) light curves available from MAST. 

% The transit signal passed all standard SPOC pipeline data validation tests, including odd-even transit comparisons, secondary-eclipse searches, bootstrap analyses, ghost diagnostics, and centroid tests. These diagnostics found no evidence for instrumental or astrophysical false-positive scenarios and showed that the transit signal is consistent with originating from TOI-2120.

\subsection{Ground-based Transit Photometry}
\begin{figure*}[!ht]
    \centering
    \includegraphics[width=\linewidth, height=\textheight, 
    keepaspectratio]{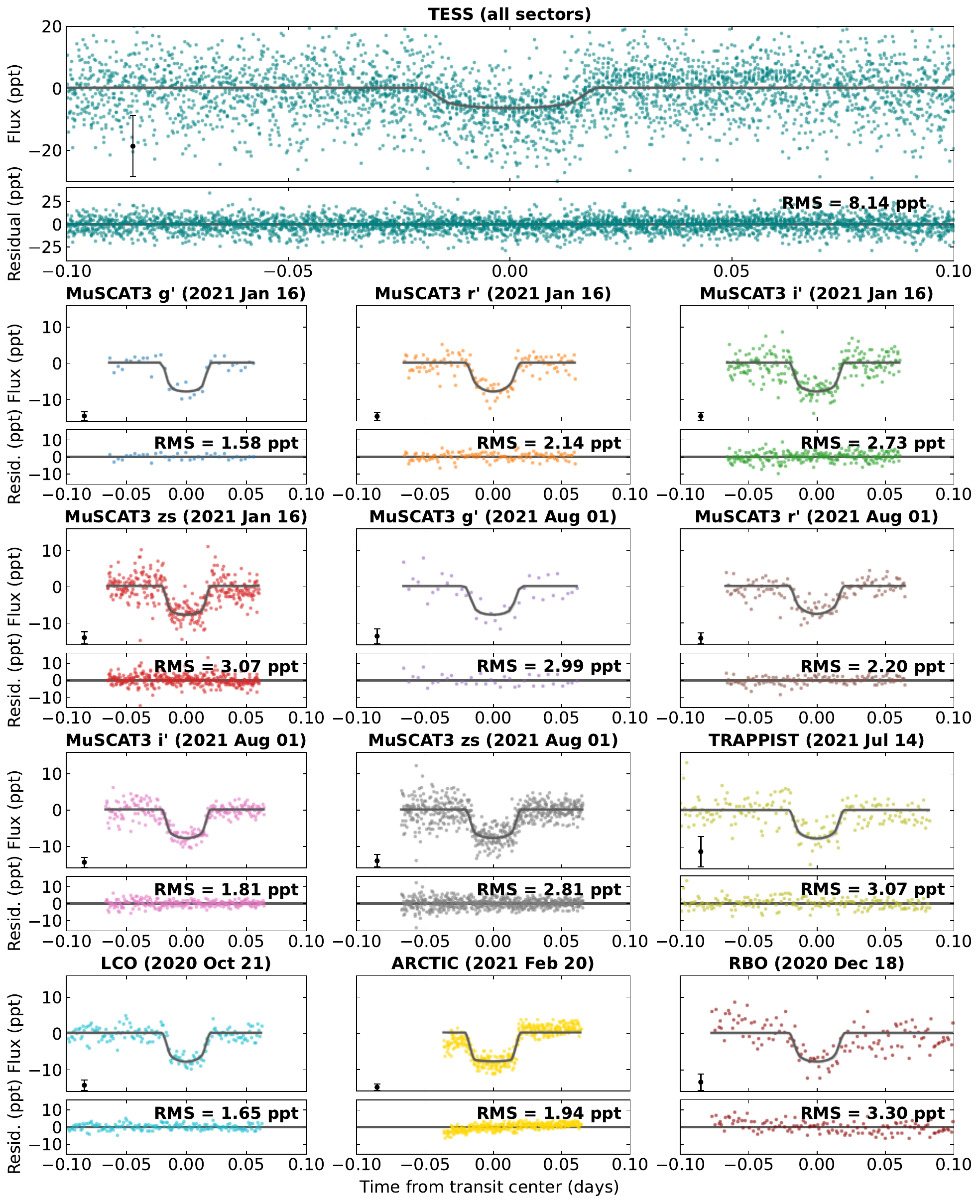}
    \caption{Phase-folded transits of TOI-2120\,b from the joint fit. The top panel shows the TESS light curve, with the residuals directly below. The panels below show the ground-based light curves with the residuals are shown directly below the light curve. The RMS scatter of the residuals is indicated in parts per thousand, and a representative photometric error bar is shown at the bottom left of each light curve.}
    \label{fig:transits}
\end{figure*}

To refine the transit parameters, we obtained two ground-based transit observations of TOI-2120\,b with the 0.6m telescope at RBO and the Astrophysical Research Consortium Telescope Imaging Camera (ARCTIC) on the ARC 3.5\,m telescope at APO. We supplement these observations with four ground-based transits reported by \citealt{Hori2024}, which were obtained with MuSCAT3 on the 2\,m Faulkes Telescope North, Sinistro on the 1\,m Las Cumbres Observatory telescope at McDonald Observatory, and TRAPPIST-North (see Figure~\ref{fig:transits}). The combined dataset provides transit coverage across multiple epochs and optical bandpasses. The observations and data reduction procedures are described below, and the complete dataset is summarized in Table~\ref{tab:gb_transits}. These observations provide higher spatial resolution than TESS and are therefore less susceptible to contamination from nearby sources.

\begin{table*}[!htp]
    \centering
    \begin{tabular}{lllllll}
    \hline
        Date (UT) & Telescope/Site - Instrument & Filter(s) & Exp. times (s) & Airmass & Ref. \\ \hline
        2020 Oct 21 & LCO 1 m/McDonald - Sinistro & Ic & 60 & 1.26-1.22-1.33 & \citealt{Hori2024} \\ 
        2020 Dec 18 & RBO 0.6 m/Red Buttes - Apogee Aspen CG47 & Bessell I & 120 & 1.20-1.97 & This work \\ 
        2021 Jan 16 & FTN 2 m/Haleakala - MuSCAT3 & g, r, i, zs & 300, 88, 40, 26 & 1.47-2.13 & \citealt{Hori2024} \\ 
        2021 Feb 20 & ARC 3.5 m/APO - ARCTIC & SDSS $i'$ & 20 & 1.36-1.85 & This work \\ 
        2021 Jul 14 & TRAPPIST-North/Oukaimeden - Andor iKon-L & z & 70 & 2.32-1.29 & \citealt{Hori2024} \\ 
        2021 Aug 1 & FTN 2 m/Haleakala - MuSCAT3 & g, r, i, zs & 300, 88, 40, 17 & 2.28-1.49 & \citealt{Hori2024} \\ \hline
    \end{tabular}
    \caption{Ground-based transit photometry of TOI-2120\,b used in this analysis}
    \label{tab:gb_transits}
\end{table*}

\subsubsection{Red Buttes Observatory (RBO) 0.6\,m}

We obtained a transit observation of TOI-2120\,b on UT 2020 December 18 using the 0.6\,m telescope at RBO (\citealt{kasper_remote_2016}). The telescope is equipped with an Apogee Aspen CG47 camera with a plate scale of 0.524\arcsecond/pixel and a field of view of 8.94\arcmin. The telescope was defocused to a FWHM of approximately 4.5\arcsecond. Observations were conducted in 2$\times$2 binning mode with 120\,s exposures through the Bessell I filter over an airmass range of 1.20–1.97.

We reduced the observations using differential aperture photometry with \texttt{AstroImageJ} (AIJ; \citealt{Collins2017}), following the methodology of \cite{stefansson2017,stefansson2018a}. Prior to photometric extraction, we applied standard bias, dark, and flat-field corrections. Aperture photometry was performed using a 7-pixel (3.7\arcsecond) radius aperture centered on the target, with inner and outer sky annuli of 12 and 17 pixels (6.3\arcsecond and 8.9\arcsecond), respectively.

\subsubsection{Astrophysical Research Consortium Telescope Imaging Camera (ARCTIC)}

We obtained a transit observation of TOI-2120\,b on UT 2021 February 20 using ARCTIC (\citealt{Huehnerhoff2016}) on the ARC 3.5\,m Telescope at APO. The observations employed the ARCTIC Engineered Diffuser, which uses a precisely patterned microlens array to transform the stellar point-spread function into a stable top-hat profile with a FWHM of approximately 8.7\arcsecond \citep{Zhao2014,stefansson2017,stefansson2018a,Stefansson2018b}. ARCTIC was operated in 4$\times$4 binning mode with the quad-amplifier fast-readout configuration, yielding a readout time of 1.3\,s. We obtained 20\,s exposures over an airmass range of 1.36–1.85.

We reduced the data using AIJ following the same procedures described above. Standard bias and flat-field corrections were applied prior to photometric extraction, but because ARCTIC dark current is negligible for exposures shorter than approximately one minute, we did not perform dark subtraction\footnote{See ARCTIC documentation: \url{https://www.apo.nmsu.edu/arc35m/Instruments/ARCTIC/\#4p3}}. Differential aperture photometry was performed using a 12-pixel (5.5\arcsecond) radius aperture centered on the target, with inner and outer sky annuli of 17 and 22 pixels (7.8\arcsecond and 10.0\arcsecond), respectively.

\subsubsection{Additional Observations from \citealt{Hori2024}}
As part of the TESS Follow-up Observing Program (TFOP; \citealt{Collins2019}), \citealt{Hori2024} obtained four ground-based transit observations of TOI-2120\,b, summarized here. 

TOI-2120\,b was observed on UT 2020 October 21 with the 1\,m Las Cumbres Observatory (LCO) Global Telescope at McDonald Observatory using the \textit{I$_{C}$} band and 60 second exposures. The telescope is equipped with a 4096$\times$4096 Sinistro camera with a pixel scale of $0\farcs389$ pixel$^{-1}$ and a $26\arcmin \times 26\arcmin$ field of view. Images were calibrated using the standard LCO Banzai pipeline \citep{McCully2018}, and differential photometry was extracted using AIJ.

Two subsequent transits were observed on UT 2021 January 16 and UT 2021 August 1 with MuSCAT3 \citep{Narita2020}, the multiband imager mounted on the 2\,m Faulkes Telescope North (FTN) operated by the LCO at Haleakala Observatory in Hawaii, USA. MuSCAT3 simultaneously images in the \textit{g}, \textit{r}, \textit{i}, and \textit{z$_{s}$} bands using a 2k$\times$2k CCD in each channel, with a pixel scale of $0\farcs27$ pixel$^{-1}$ and a $9\farcm1 \times 9\farcm1$ field of view. Exposure times were 300, 88, 40, and 17 seconds on UT 2021 January 16 and 300, 88, 40, and 17 seconds on UT 2021 August 1. Image calibration and aperture photometry were performed using the \citealt{Fukui2011} pipeline.

The fourth transit was observed on UT 2021 July 14 with the 0.6\,m TRAPPIST-North telescope at the Ouka"{i}meden Observatory in Morocco \citep{Jehin2011,Gillon2013,Barkaoui2019}. TRAPPIST-North is equipped with a thermoelectrically cooled 2k$\times$2k Andor iKon-L BEX2-DD CCD with a pixel scale of $0\farcs64$ pixel$^{-1}$ and a $20\arcmin \times 20\arcmin$ field of view. The observation was obtained in the Sloan \textit{z'} band with 70 second exposures, and image reduction and differential photometry were performed using \texttt{prose} \citep{Garcia2022}.

\subsection{Radial Velocities}
We use high-precision RV measurements to constrain the mass of TOI-2120\,b and characterize its orbit. Here, we make use of new observations from the Habitable-zone Planet Finder (HPF), together with the Subaru/IRD dataset reported by \citealt{Hori2024}.

\begin{table}[!htbp]
    \centering
    \begin{tabular}{ccc}
    \hline
        BJD\textsubscript{TDB}(days) & RV(m s\textsuperscript{-1}) & $\sigma$ (m s\textsuperscript{-1}) \\ \hline
        2459092.88670 & -12.28 & 7.51 \\ 
        2459094.89132 & -1.69 & 8.31 \\ 
        2459123.85286 & -33.36 & 11.81 \\ 
        2459125.82608 & -8.84 & 9.13 \\ 
        2459126.82480 & -7.83 & 12.49 \\ 
        2459127.81895 & -11.26 & 7.58 \\ 
        2459128.81729 & -12.42 & 8.47 \\ 
        2459130.81821 & 11.23 & 7.79 \\ 
        2459132.80459 & 0.07 & 6.79 \\ 
        2459147.76401 & 1.57 & 11.01 \\ 
        2459186.66122 & -7.56 & 25.15 \\ 
        2459192.64518 & -15.72 & 7.18 \\ 
        2459193.63685 & -16.55 & 7.06 \\ 
        2459204.62145 & -7.69 & 10.95 \\ 
        2459205.60154 & 9.89 & 7.97 \\ 
        2459206.59565 & 14.88 & 17.79 \\ 
        2459210.62026 & 2.57 & 13.89 \\ 
        2459211.60638 & -1.07 & 7.47 \\ 
        2459212.62295 & 15.67 & 11.34 \\ 
        2459219.60572 & 9.51 & 8.30 \\ 
        2459221.57510 & 0.68 & 8.33 \\ 
        2459227.56524 & 2.20 & 10.28 \\ 
        2459433.96423 & -23.05 & 9.67 \\ 
        2459436.97560 & 6.19 & 8.24 \\ 
        2459445.94811 & -16.69 & 11.40 \\ 
        2459454.91949 & -5.02 & 8.95 \\ 
        2459458.89998 & -7.84 & 11.29 \\ 
        2459460.90632 & -2.05 & 8.23 \\ 
        2459490.82920 & -20.30 & 10.03 \\ 
        2459503.81899 & -26.58 & 12.31 \\ 
        2459504.81221 & 3.79 & 12.65 \\ 
        2459509.77263 & -0.59 & 9.48 \\ 
        2459511.81167 & -5.97 & 15.74 \\ 
        2459512.81161 & -6.71 & 11.29 \\ 
        2459534.72649 & -10.87 & 13.65 \\ 
        2459535.73161 & -8.79 & 9.58 \\ 
        2459560.63776 & 10.64 & 9.49 \\ 
        2459562.62722 & -27.07 & 10.51 \\ \hline
    \end{tabular}
    \caption{RV observations of TOI-2120, taken with HPF, binned down to one day.}
    \label{tab:hpf_rvs}
\end{table}

\subsubsection{Habitable-zone Planet Finder (HPF)}
HPF (\citealt{mahadevan_habitable-zone_2012,mahadevan_habitable-zone_2014}) is a thermally stabilized, fiber-fed near-infrared spectrograph \citep{kanodia_overview_2018,stefansson2016} installed on the 10\,m Hobby-Eberly Telescope (HET; \citealt{ramsey_early_1998,Hill2021}) at McDonald Observatory. HPF covers wavelengths from 808–1278\,nm with a resolving power of R$\sim$50,000 and has demonstrated an on-sky instrumental stability of approximately 1\,m s$^-1$ \citep{metcalf2019,lubin2021}. HET employs a fixed-altitude, roving-pupil design and operates as a fully queue-scheduled facility, with observations carried out by resident astronomers \citep{shetrone2007}.

\begin{figure*}[!ht]
    \centering
    \includegraphics[width=\linewidth]{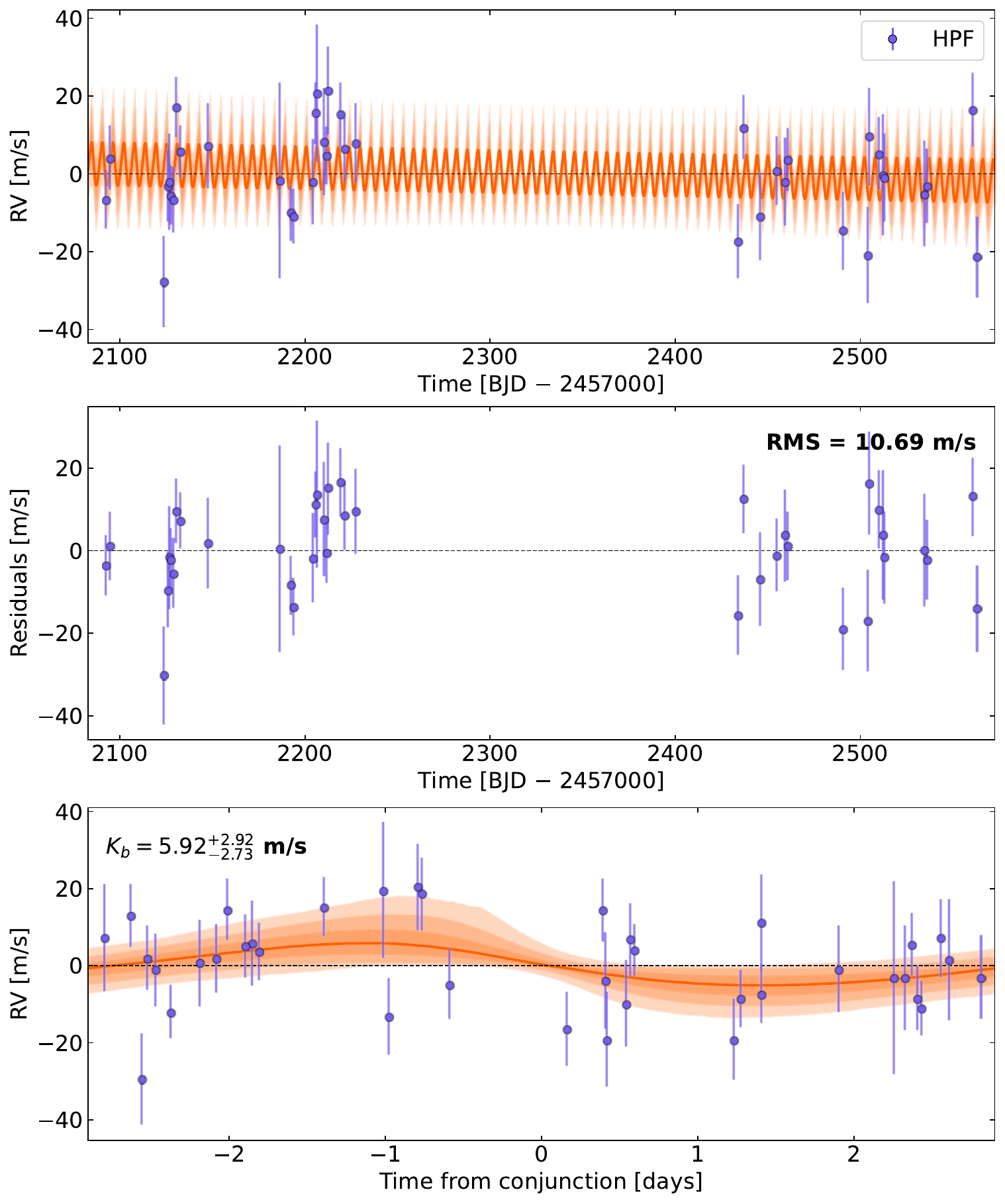}
    \caption{HPF RVs of TOI-2120 (purple points) and the best-fit orbit from the joint transit–RV analysis. \textbf{Top:} RV time series after removing the RV zero point. The orange curve shows the median posterior model, including the Keplerian orbit and fitted linear trend, with dark-to-light shading indicating the 1, 2, and 3$\sigma$ credible intervals. \textbf{Middle:} Residuals about the median model, with an RMS scatter of 10.69\,m\,s$^{-1}$. \textbf{Bottom:} RVs phase-folded on the orbital period ($P = 5.79983$ d), together with the folded Keplerian model and corresponding 1, 2, and 3$\sigma$ credible intervals.}
    \label{fig:rvs}
\end{figure*}

We observed TOI-2120 with HPF over 38 visits spanning UT 2020 August 31 to 2021 December 14. Each visit consisted of three 649\,s exposures, except for three visits that were limited to two exposures each because of deteriorating observing conditions. The individual exposures from each visit were combined to derive a single RV measurement for each observing epoch. The observations reached a median signal-to-noise ratio (S/N) per pixel of 52 per exposure, measured in HPF order index 57 near $\sim$1000\,nm. The resulting RV measurements are shown both raw and phase-folded in Figure~\ref{fig:rvs}.

To avoid contamination of the science spectra from scattered light, we did not employ the simultaneous calibration mode of HPF, which uses a near-infrared laser frequency comb (LFC; \citealt{metcalf2019}) to illuminate a calibration fiber alongside the science fiber. Instead, wavelength solutions were determined by interpolating between LFC calibration exposures acquired throughout the night. This procedure has been demonstrated to provide wavelength calibration and instrumental drift corrections at the $\sim$30\,cm s$^{-1}$ level \citep{stefansson_sub-neptune-sized_2020}.

We reduced the HPF observations using the HPF data reduction pipeline. The raw detector images were processed with \texttt{HxRGproc} \citep{ninan_habitable-zone_2018}, and RVs were extracted using a modified version of the SpEctrum Radial Velocity AnaLyser (\texttt{SERVAL}; \citealt{zechmeister_spectrum_2018, stefansson_sub-neptune-sized_2020, Stefansson2023}) following the template-matching approach of \cite{anglada_escude2012}. We constructed the stellar template from all available TOI-2120 spectra after masking telluric regions identified with a synthetic telluric mask generated using \texttt{telfit} \citep{gullikson2014}, a Python wrapper for the Line-by-Line Radiative Transfer Model \citep{clough2005atmospheric}. Barycentric corrections were calculated with \texttt{barycorrpy} \citep{kanodia_wright_2018}, which implements the algorithms described by \cite{wright_eastman_2014}. The HPF RVs used in this analysis are reported in Table~\ref{tab:hpf_rvs}.

\subsubsection{Subaru/IRD}
\citealt{Hori2024} obtained 24 high-resolution spectra of TOI-2120 with the InfraRed Doppler (IRD) instrument on the 8.2\,m Subaru Telescope \citep{Tamura2012,Kotani2018} as part of the Subaru/IRD TFOP program. IRD is a fiber-fed near-IR spectrograph covering 930--1740\,nm at a spectral resolution of $R\sim70,000$. They reduced these spectra following \citealt{Kuzuhara2018} and \citealt{Hirano2020} and discarded observations with S/N$<$18 at 1000 nm, as well as those affected by clouds or instrumental systematics, leaving 17 spectra with S/N of 30--60 pixel$^{-1}$. RVs were extracted using the forward modeling procedure in \citealt{Hirano2020}, which accounts for telluric absorption and time-dependent variations in the instrument profile. 

\subsection{High-Contrast Imaging with ShaneAO}
High-angular resolution imaging was obtained to search for nearby stellar companions and background sources that could contaminate the transit photometry or produce astrophysical false positive scenarios. Previous observations of TOI-2120 included speckle imaging with the SAI-2.5\,m telescope in the $I$ band and Gemini-N/'Alopeke in the 562 and 832\,nm bands, which found no nearby stellar companions within 2\arcsec \citep{Hori2024}. To complement these observations and extend the wavelength coverage into the near-infrared, we obtained adaptive optics (AO) observations of TOI-2120 with the Shane Adaptive optics infraRed Camera and Spectrograph (ShARCS; \citealt{srinath2014}) on the 3\,m Shane Telescope at Lick Observatory on Mt. Hamilton, California. Observations were obtained in the $K_s$ band on UT 2021 February 27 using natural guide star mode. Although observing conditions were relatively poor, with seeing of approximately 2\arcsecond, the observations were stable throughout the sequence and provide an additional constraint on potential red stellar companions.

% we observed TOI-2120 with the adaptive optics (AO) system on the 3\,m Shane Telescope at Lick Observatory on Mt.~Hamilton, California. Observations were obtained using the Shane Adaptive optics infraRed Camera and Spectrograph (ShARCS; \citealt{srinath2014}) in the K$_s$ band on UT 2021 February 27. The observations were conducted in natural guide star mode under relatively poor but stable seeing conditions of approximately 2\arcsecond.

\begin{figure}[ht]
    \centering
    \includegraphics[width=\columnwidth]{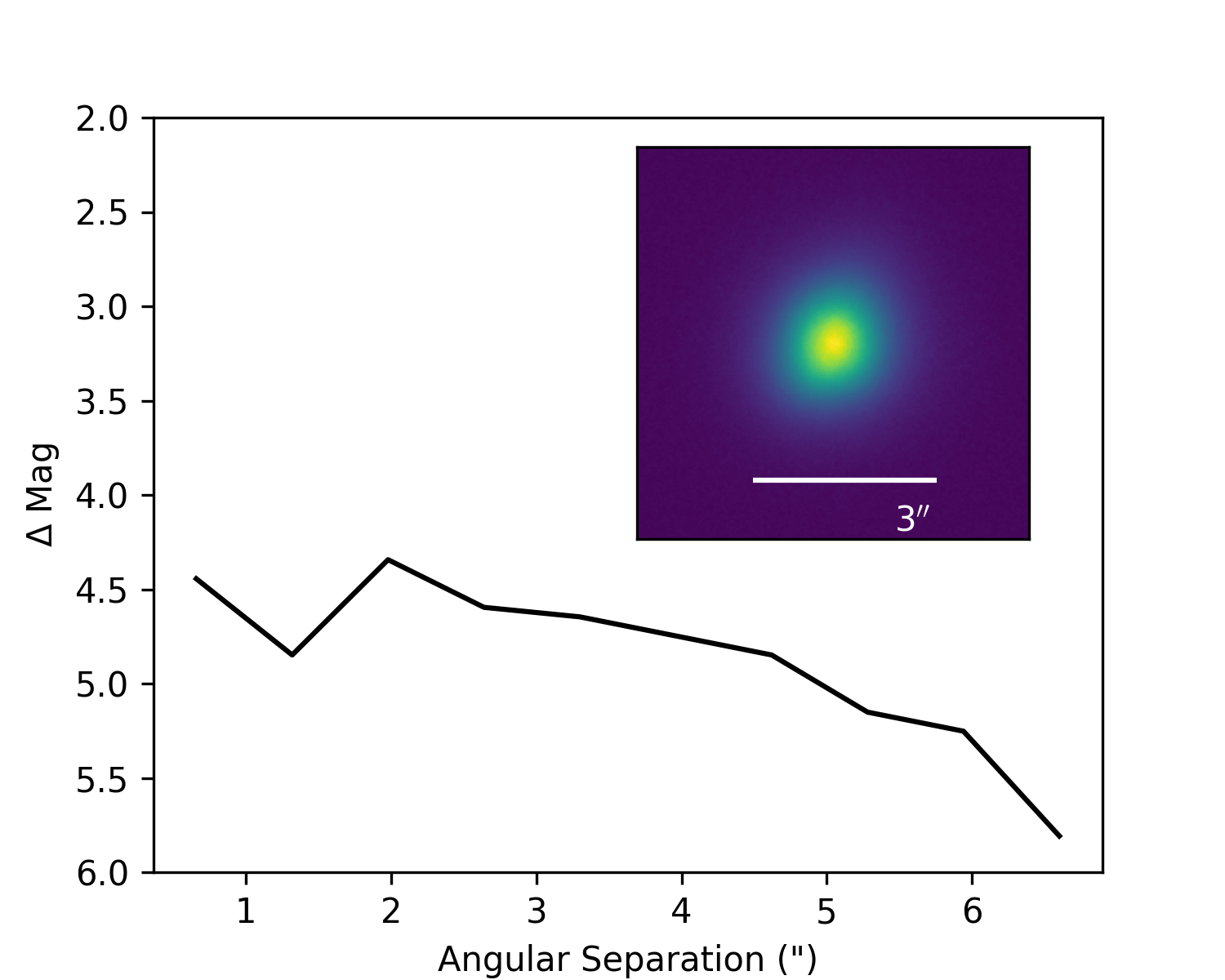}
    \caption{Ks-band contrast curve and ShaneAO image (inset) of TOI-2120. High-contrast imaging rules out companions with $\Delta$K$_{s} < $4.5 at separations greater than 1\arcsecond.}
    \label{fig:shaneao}
\end{figure}

We used a five-point dither pattern \citep{furlan2017} to mitigate detector systematics and improve estimation of the sky background. A total open-shutter exposure time of 2060\,s was obtained across eleven individual exposures. The data were reduced using a custom pipeline that performed image calibration, registration, and combination. We generated a 5$\sigma$ contrast curve following the methodology described by \cite{espinoza2016}. The resulting AO image and contrast curve are shown in Figure~\ref{fig:shaneao}. Although the observing conditions limited our sensitivity to very close companions, the observations exclude companions with $\Delta$K$_{s} < $4.5 beyond 1\arcsecond, ruling out the majority of background eclipsing-binary false-positive scenarios.

% \subsubsection{Speckle Imaging}
% Additional high-resolution imaging observations were reported by \cite{Hori2024}. TOI-2120 was observed with the Speckle Polarimeter mounted on the 2.5 m SAI telescope and with the 'Alopeke speckle imager on the Gemini North telescope. These observations provide diffraction-limited imaging at optical wavelengths and are sensitive to close stellar companions that may not be detectable in seeing-limited observations. No nearby stellar companions were detected within the sensitivity limits of either dataset. The resulting contrast curves place additional constraints on unresolved sources within a few arcseconds of the target star and further reduce the likelihood of background eclipsing binaries or bound stellar companions capable of producing the observed transit signal. We include these observations for completeness and as additional validation of the planetary interpretation; however, they are not incorporated into the analysis presented in this work.

\section{Stellar Activity}\label{sec:activity}
We used the HPF spectra to assess whether stellar activity could contribute to the observed RV variations. To search for activity-related signals, we examined several commonly used near-infrared activity indicators available within the HPF wavelength range, including the differential line width (dLW), chromatic index (CRX; \citealt{zechmeister_spectrum_2018}), and the calcium infrared triplet (Ca IRT) indices at air wavelengths of 8498, 8542, and 8662\AA.

None of the activity indicators showed a significant correlation with the RVs. We quantified these correlations using the Kendall rank correlation coefficient ($\tau$; \citealt{kendall1948rank}), which ranged from 0.01 to 0.09 across all indicators. The generalized Lomb–Scargle (GLS) periodograms and corner plots for the HPF RVs and activity indicators are presented in Appendix~\ref{app:activity}.

We additionally searched the activity indicators for signatures of stellar rotation but found no significant periodicity in the GLS periodograms that could be associated with rotationally modulated stellar activity. We therefore cannot directly constrain the stellar rotation period from either the RV measurements or the activity indicators. The strongest constraint on the rotation rate comes from the projected rotational velocity, $v\sin i<$2\,km s$^{-1}$, inferred from the absence of detectable rotational broadening in the HPF spectra.
Taken together, the lack of significant correlations between the RVs and activity indicators, along with the low projected rotational velocity, suggests that TOI-2120 is a relatively inactive M dwarf. Stellar activity is therefore unlikely to be a significant contributor to the HPF RV measurements.

\begin{deluxetable*}{lccc}
\tablecaption{Summary of stellar parameters for TOI-2120. \label{tab:stellarparam}}
\tablehead{\colhead{~~~Parameter}&  \colhead{Description}&
\colhead{Value}&
\colhead{Reference}}
\startdata
\multicolumn{4}{l}{\hspace{-0.2cm} Main identifiers:}  \\
~~~TOI & \tess Object of Interest & 2120 & \tess mission \\
~~~TIC & \tess Input Catalog  & 389900760 & Stassun 2018 \\
~~~2MASS & $\cdots$ & J01340099+6530537 & 2MASS  \\
~~~Gaia DR3 & $\cdots$ & 513299860904522752  & Gaia DR3\\
\multicolumn{4}{l}{\hspace{-0.2cm} Equatorial Coordinates, Proper Motion and Spectral Type:} \\
~~~$\alpha_{\mathrm{ICRS, 2016.0}}$ &  Right Ascension (RA) & 01:34:01.76 & Gaia DR3\\
~~~$\delta_{\mathrm{ICRS, 2016.0}}$ &  Declination (Dec) & +65:30:50.36 & Gaia DR3\\
~~~$\mu_{\alpha}$ &  Proper motion (RA, \unit{mas/yr}) &  306.282 & Gaia DR3\\
~~~$\mu_{\delta}$ &  Proper motion (Dec, \unit{mas/yr}) & -215.531 & Gaia DR3 \\
~~~$d$ &  Distance in pc  & $32.176\pm0.065$ & Bailer-Jones \\
% ~~~$A_{V,max}$ & Maximum visual extinction & \rachel{xx}  & Green\\
\multicolumn{4}{l}{\hspace{-0.2cm} Optical and near-infrared magnitudes:}  \\
~~~$B$ & Johnson B mag & $17.241\pm0.464$ & APASS\\
~~~$V$ & Johnson V mag & $15.648\pm0.105$ & APASS\\
~~~$T$  & \tess magnitude & $12.823\pm0.007$  & Stassun 2018 \\
~~~$J$ & $J$ mag & $11.1\pm0.023$ & 2MASS\\
~~~$H$ & $H$ mag & $10.452\pm0.026$ & 2MASS\\
~~~$K_s$ & $K_s$ mag & $10.205\pm0.023$ & 2MASS\\
~~~$G$ & $G$ mag & $14.159\pm0.001$ & Gaia DR3\\
\multicolumn{4}{l}{\hspace{-0.2cm} Stellar Parameters:}\\
~~~$T_{\mathrm{eff}}$ &  Effective temperature in \unit{K} & $3138\pm50$ & Gore 2024\\
~~~$\mathrm{[Fe/H]}$ &  Metallicity in dex & $0.128\pm0.085$ & Gore 2024\\
~~~$\log(g)$ & Surface gravity in cgs units & $5.00\pm0.05$ & Derived\\
~~~$M_*$ &  Mass in $M_{\odot}$ & $0.211\pm0.020$ & Gore 2024\\
~~~$R_*$ &  Radius in $R_{\odot}$ & $0.240\pm0.007$ & Gore 2024\\
~~~$L_*$ &  Luminosity in $L_{\odot}$ & $0.0050\pm0.0001$ & Gore 2024\\
~~~$\rho_*$ &  Density in $\unit{g/cm^{3}}$ & $21.52\pm2.78$ & Derived\\
% ~~~Age & Age in Gyrs & $9.7^{+3.9}_{-2.7}$ & This work\\
% ~~~$A_v$ & Visual extinction in mag & $0.075^{+0.074}_{-0.053}$ & This work\\
\multicolumn{4}{l}{\hspace{-0.2cm} Other Stellar Parameters:}           \\
~~~\vsini &  Rotational velocity in \unit{km/s}  & $<$2 & This work\\
~~~$\Delta$RV &  ``Absolute'' radial velocity in \unit{km/s} & 21.37  & Gaia DR3\\
% ~~~$U, V, W$ &  Galactic velocities in \unit{km/s} &  \rachel{xx} & This work\\
\enddata
\tablenotetext{}{References are: Stassun \citep{stassun2018_tic}, 2MASS \citep{cutri2003_2mass}, Gaia DR3 \citep{Gaiadr3}, Bailer-Jones \citep{bailerjones2021_gdr3}, APASS \citep{henden2018_apass}, Gore 2024 \citep{Gore2024}}
\end{deluxetable*}

\section{Joint Fitting} \label{sec:joint}

We performed a joint fit to the photometry and the RVs using the \texttt{exoplanet} package \citep{exoplanet_dfm}, which is based on \texttt{PyMC3}, a Hamiltonian Monte Carlo package \citep{salvatier2016probabilistic}. The \texttt{exoplanet} package uses \texttt{starry} \citep{luger2019,agol2020} to simulate planetary transits, employing the analytical transit models by \citet{mandel_agol2002} along with a quadratic limb-darkening law. Limb-darkening priors in \texttt{exoplanet} are based on the reparameterization by \citet{kipping2023a} for uninformative sampling. Stellar parameters for the fit were adopted from \citet{Gore2024} and are reported in Table~\ref{tab:stellarparam}.

Each transit was fitted independently, with separate limb-darkening coefficients for each transit. All TESS sectors were treated as a single instrument, whereas ground-based observations from different facilities or observing epochs were treated as separate instruments. Because of the large 21\arcsecond TESS pixels, the PDCSAP light curves include a dilution correction to account for photometric contamination from nearby sources. We nevertheless allow the TESS dilution factor to vary in the joint fit to account for uncertainties in this correction. In contrast, the ground-based transit observations have substantially narrower point-spread functions and therefore reduced contamination from nearby sources. We consequently fix the dilution factor to unity for all ground-based datasets.

We modeled the RVs using a standard Keplerian model with free \textit{e} and $\omega$, while accounting for independent RV offsets, jitter terms, and linear RV slopes for each RV dataset. We include the stellar mass, radius, and temperature, as well as the transit midpoint, orbital period, and transit depth, as priors in our posterior fits. We run four chains with 1500 burn-in steps followed by an additional 4500 steps to ensure robust estimation of the planetary parameters. To assess whether the posterior distributions are well mixed and independent, we evaluate convergence using the Gelman-Rubin statistic, for which $\hat{R} \sim 1$ indicates strong convergence \citep{Ford2006}. The fitted planetary parameters are listed in Table~\ref{tab:orbitalparam}.

We first fit the transit photometry jointly with both the HPF and IRD RVs. We find a large amplitude linear trend in the IRD RVs of $-16\pm6$ m s$^{-1}$ yr$^{-1}$, while the HPF RVs show no significant trend ($-2.7\pm3.5$ m s$^{-1}$ yr$^{-1}$). The slope in the IRD RVs is consistent with what \citealt{Hori2024} finds, although they do not discuss the cause of this trend in their analysis. The presence of this trend, together with the limited number of IRD RVs ($\sim15$), results in a low-precision constraint on the planetary RV signal (see Appendix~\ref{app:ird}). The lower precision and the trend may also be due to additional systematics in the IRD RVs that are not captured in the published measurements. Because we do not have access to the underlying IRD data or the details of the IRD reduction and systematic-error modeling, we are unable to further investigate the origin of this trend or determine whether additional modeling could improve the RV constraints. We therefore exclude the IRD velocities from our final fit and use the HPF data alone, for which the instrumental drifts are corrected for.

Using just the HPF RVs with the ground-based transit photometry, we perform two fits, one with $e,\omega$ fixed and one in which both parameters are allowed to vary. We find consistent ($\sim3\sigma$) planetary properties between the two fits, and therefore adopt the free $e,\omega$ solution for our analysis. We find TOI-2120\,b to be a sub-Neptune with a radius of $2.306^{+0.075}_{-0.073}$\Rearth and a mass of $5.81^{+2.94}_{-2.69}$\Mearth, corresponding to a bulk density of $2.62^{+1.37}_{-1.22}$ g cm$^{-3}$. A summary of the fitted planetary parameters, their corresponding confidence intervals and how they compare with \citet{Hori2024} is presented in Table~\ref{tab:orbitalparam}.

\begin{deluxetable*}{llcc}  %{lp{3cm}c}
\tablecaption{Summary of orbital and physical parameters for TOI-2120 b. \label{tab:orbitalparam}}
\tablehead{\colhead{} & \colhead{} & \multicolumn{2}{c}{Value} \\
\cline{3-4}
\colhead{~~~Parameter} & \colhead{Units} & \colhead{This work\textsuperscript{a}} & \colhead{Hori et al. (2024)\textsuperscript{b}}}
\startdata
\sidehead{Orbital Parameters:}
Orbital Period & P (days) & $5.79982836^{+0.00000537}_{-0.00000535}$ & $5.7998164\pm0.0000035$\\
Eccentricity & e & $0.140^{+0.129}_{-0.087}$ & $0.32\pm0.08$\\
Argument of Periastron & $\omega$ (radians) & $0.6106^{+1.515}_{-0.671}$ & \nodata\\
Semi-amplitude Velocity & K (m s$^{-1}$) & $5.9^{+2.9}_{-2.7}$ (HPF) & $<7.3$\textsuperscript{f} (IRD)\\
Systemic Velocity & $\gamma$ (m s$^{-1}$) & $-5.5^{+1.9}_{-1.8}$\textsuperscript{c} (HPF) & \nodata\\
RV trend & $\frac{dv}{dt}$ (m s$^{-1}$ yr$^{-1}$) & $-3.33^{+3.87}_{-4.00}$ (HPF) & $-16.07\pm2.56$\textsuperscript{g} (IRD)\\
RV jitter & $\sigma$ (m s$^{-1}$) & $5.11^{+2.67}_{-2.80}$ (HPF) & $3.5\pm2.0$ (IRD)\\
\hline
\sidehead{Transit Parameters:}
Transit Midpoint & T\textsubscript{0} (BJD\textsubscript{TDB}) & $2458795.82201^{+0.00047}_{-0.00046}$ & $2458795.82368\pm0.00041$\\
Impact parameter & b & $0.74^{+0.02}_{-0.03}$ & $0.55^{+0.07}_{-0.10}$\\
Scaled Radius & $R_p/R_*$ & $0.088^{+0.001}_{-0.001}$ & $0.0796\pm0.0014$\\
Scaled Semi-major Axis & $a/R_*$ & $34.06^{+1.41}_{-1.40}$ & $33.21^{+0.87}_{-0.83}$\\
Orbital Inclination & i (degrees) & $88.64^{+0.12}_{-0.14}$ & $88.67^{+0.14}_{-0.11}$\\
Transit Duration & T\textsubscript{14} (days) & $0.043^{+0.002}_{-0.002}$ & \nodata\\
Photometric Jitter\textsuperscript{d} & $\sigma_{\mathrm{TESS}}$ (ppm) & $1468.23^{+38.30}_{-38.22}$ & \nodata\\
 & $\sigma_{\mathrm{ARCTIC}}$ (ppm) & $1644.62^{+85.23}_{-81.96}$ & \nodata\\
 & $\sigma_{\mathrm{RBO}}$ (ppm) & $2446.34^{+303.90}_{-290.68}$ & \nodata\\
 & $\sigma_{\mathrm{TRAPPIST}}$ (ppm) & $22.29^{+165.45}_{-19.57}$ & \nodata\\
 & $\sigma_{\mathrm{LCO}}$ (ppm) & $648.30^{+261.34}_{-599.64}$ & \nodata\\
~~~MuSCAT3 (2021 Jan 16) & $\sigma_{g'}$ (ppm) & $705.70^{+452.16}_{-687.64}$ & \nodata\\
 & $\sigma_{r'}$ (ppm) & $1840.25^{+177.40}_{-165.94}$ & \nodata\\
 & $\sigma_{i'}$ (ppm) & $2472.53^{+143.93}_{-132.07}$ & \nodata\\
 & $\sigma_{z_s}$ (ppm) & $2563.94^{+144.47}_{-135.00}$ & \nodata\\
~~~MuSCAT3 (2021 Aug 1) & $\sigma_{g'}$ (ppm) & $2143.98^{+516.87}_{-498.06}$ & \nodata\\
 & $\sigma_{r'}$ (ppm) & $1490.25^{+207.19}_{-198.95}$ & \nodata\\
 & $\sigma_{i'}$ (ppm) & $1124.58^{+131.06}_{-129.80}$ & \nodata\\
 & $\sigma_{z_s}$ (ppm) & $2110.51^{+110.84}_{-107.22}$ & \nodata\\
Dilution\textsuperscript{e} & $D_{\mathrm{TESS}}$ & $0.833^{+0.043}_{-0.037}$ & \nodata\\
\hline
\sidehead{Planetary Parameters:} 
Mass & $M_p$ (\Mearth) & $5.8^{+2.9}_{-2.7}$ & $<6.8$\textsuperscript{f}\\
Radius & $R_p$ (\Rearth) & $2.31^{+0.07}_{-0.07}$ & $2.122\pm0.065$\\
Density & $\rho_p$ (g cm$^{-3}$) & $2.616^{+1.374}_{-1.220}$ & \nodata\\
Semimajor Axis & a (au) & $0.0379^{+0.0011}_{-0.0012}$ & $0.0377\pm0.0013$\\
Planetary Insolation & S (S$_{\oplus}$) & $3.47^{+0.36}_{-0.36}$ & $3.61^{+0.24}_{-0.22}$\\
Equilibrium Temperature\textsuperscript{h} & $T_{eq}$ (K) & $380\pm10$ & $384\pm6$\\
\enddata
\tablecomments{\textsuperscript{a}Values are the 16th, 50th, and 84th percentiles of the posteriors.
\textsuperscript{b}Median and 1$\sigma$ bounds of the posteriors from Hori et al. (2024, Tables 5 and 6).
\textsuperscript{c}In addition to the absolute RV given in Table~\ref{tab:stellarparam}.
\textsuperscript{d}Per-observation jitter added in quadrature to the photometric instrument errors.
\textsuperscript{e}Uncorrected dilution due to background stars in the TESS aperture.
\textsuperscript{f}2$\sigma$ upper limit.
\textsuperscript{g}Converted from m\,s$^{-1}$\,day$^{-1}$.
\textsuperscript{h}Assuming a Bond albedo of 0.}
\end{deluxetable*}

\section{Discussion}\label{sec:discussion}
\subsection{Comparison with \citealt{Hori2024}}
\citet{Hori2024} used TESS along with multicolor ground-based transit photometry to identify TOI-2120\,b as a transiting $2.12\pm0.065$\Rearth sub-Neptune and placed a $2\sigma$ upper limit of 6.8\Mearth on its mass using Subaru/IRD RVs. They also found the planet to be moderately eccentric ($e=0.32\pm0.08$). In our analysis, we revisit the planetary parameters using the same transit data as in \citet{Hori2024}, along with additional transits from ARCTIC and RBO, as well as new RVs from HPF. We adopt the stellar parameters for TOI-2120 reported by \citet{Gore2024}. We note that \citet{Hori2024} independently derived stellar parameters primarily from photometric relations (Section 3.1.2), and that some of the differences between our reported planetary parameters and theirs therefore arise from differences in the adopted stellar parameters rather than from the planetary fits themselves.

% As in \citet{Hori2024}, we also adopt the stellar parameters for TOI-2120 from \citet{Gore2024}, allowing us to compare the planetary parameters without introducing differences in the underlying stellar parameters.

We find a planetary radius of $2.306^{+0.075}_{-0.073}$\,\Rearth, compared to $2.12\pm0.065$\,\Rearth from \citet{Hori2024}. Although our fit favors a larger planet, the two measurements are consistent within $2\sigma$. We also find a higher impact parameter of $b=0.742^{+0.024}_{-0.029}$ compared to \citealt{Hori2024}'s $b=0.55^{+0.07}_{-0.10}$. We attribute the difference in eccentricity which could influence the resulting impact parameter, and therefore planet radius, to the additional ARCTIC and RBO transits.

By using HPF RVs, we are able to better constrain the TOI-2120 mass of $5.81^{+2.94}_{-2.69}$\Mearth to $2\sigma$, which is consistent with the upper limit of $<6.8$\,\Mearth measured by \citealt{Hori2024} using IRD RVs. Combining this with our measured planetary radius, we get a density of $2.62^{+1.37}_{-1.22}$ g cm$^{-3}$, which is consistent with a low-density sub-Neptune. However, we note that the uncertainty here is dominated by the relatively weak RV mass constraint, and therefore we do not use the density to make any detailed compositional inferences.

Our eccentricity constraint is less definitive. We find $e=0.140^{+0.129}_{-0.087}$, which, when compared to \citealt{Hori2024}'s $e=0.32\pm0.08$, is consistent within $2\sigma$. However, our fit only has a $1.6\sigma$ preference for a non-zero eccentricity. We therefore cannot distinguish between a modestly eccentric and circular orbit, and this may suggest that the orbital eccentricity remains sensitive to the choice of RV dataset.

\begin{figure*}[!ht]
    \centering
    \includegraphics[width=\linewidth]{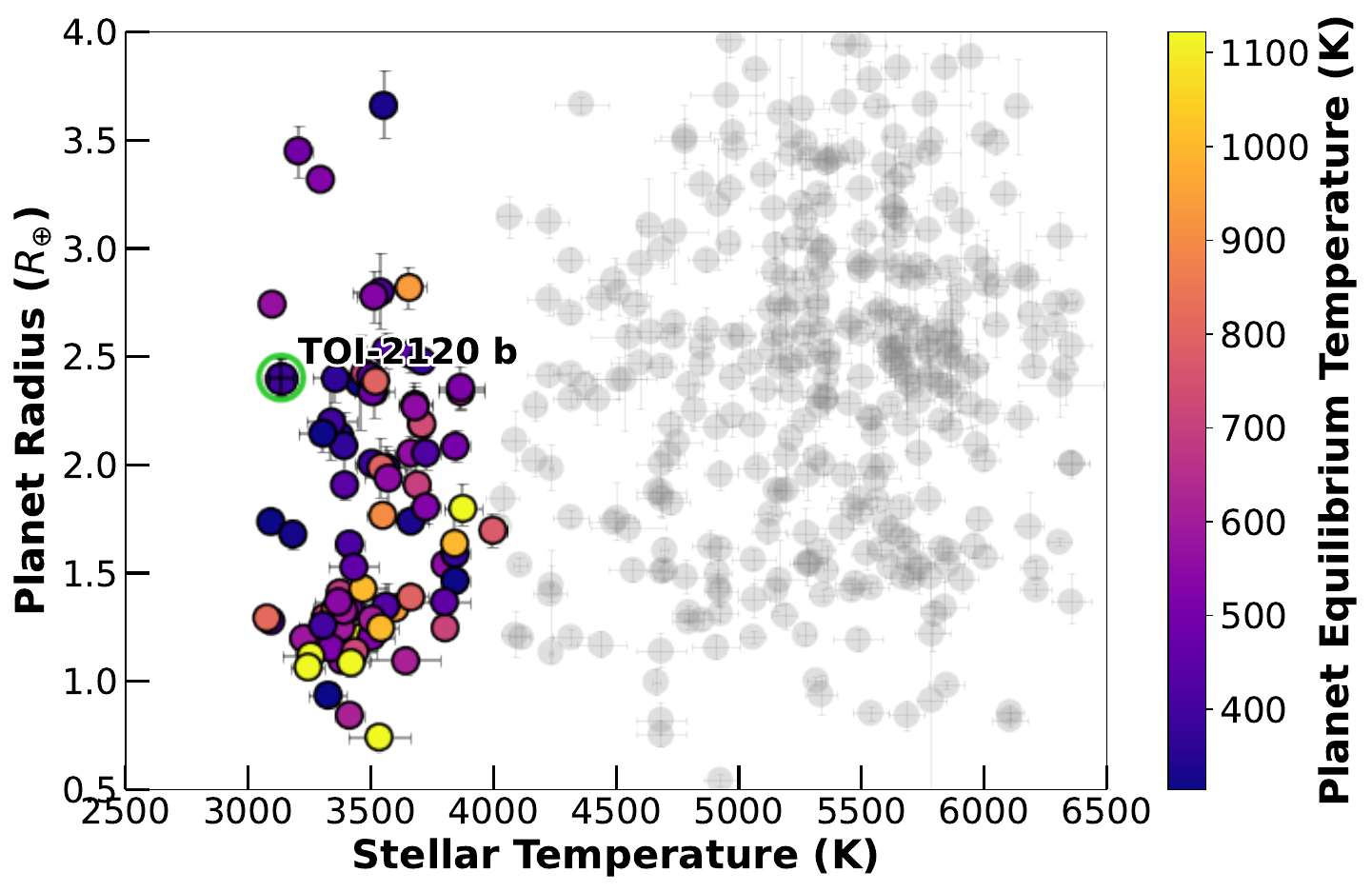}
    \caption{Planet Radius vs. Stellar Temperature for TOI-2120\,b (highlighted by a green circle) alongside other planets around M dwarfs with radii $<$4\,\Rearth. The planet equilibrium temperature is color-coded for each planet, and for comparison, planets around FGK-type stars are displayed in gray.}
    \label{fig:coolsystem}
\end{figure*}

\subsection{TOI-2120\,b in Context with Small Planets around M dwarfs}
To place TOI-2120,b in the context of the small planet population around M dwarfs, we compiled a sample of transiting planets smaller than 4\,\Rearth with measured masses and radii from the Planetary Systems Composite Parameters table \citep{PSCompPars} of the NASA Exoplanet Archive \citep[NEA;][]{Christiansen2025}, accessed on 2026 October.

TOI-2120\,b is particularly relevant to the emerging demographics of planets around the lowest-mass stars. With a mass of 0.211\,\Msun and $T_{\rm eff}=3100$\,K, TOI-2120 lies toward the mid-to-late M dwarf regime (see Figure~\ref{fig:coolsystem}), where the observed small-planet population appears to differ from that around FGK and early-M stars. Studies of earlier M dwarfs have identified a radius valley, although its dependence on incident flux remains disputed: \citet{cloutiermenou2020} measured a slope opposite to that expected from thermally driven atmospheric loss, whereas subsequent analyses have found slopes ranging from nearly flat to consistent with the negative slope measured around FGK stars \citep[e.g.,][]{vaneylen2021,luquepalle2022}. More recently, \citet{gillis2026} measured a unimodal radius distribution peaked at $1.25\pm0.05$\,\Rearth for 77 vetted TESS planet candidates around mid-to-late M dwarfs, with super-Earths outnumbering sub-Neptunes by $\sim$5.5:1 within 30,days ($0.954\pm0.147$ versus $0.148\pm0.045$ planets per star). They interpret the absence of a bimodal distribution, driven by this paucity of sub-Neptunes, as the disappearance of the radius valley around the lowest-mass stars. The presence of a $>$2\,\Rearth planet around a $3100$\,K star therefore places TOI-2120\,b in a sparsely populated region of the emerging mid-to-late M dwarf planet distribution and provides an individual system with which to probe the transition between the planet populations observed around earlier and later M dwarfs.

\begin{figure*}[!ht]
    \centering
    \includegraphics[width=\linewidth]{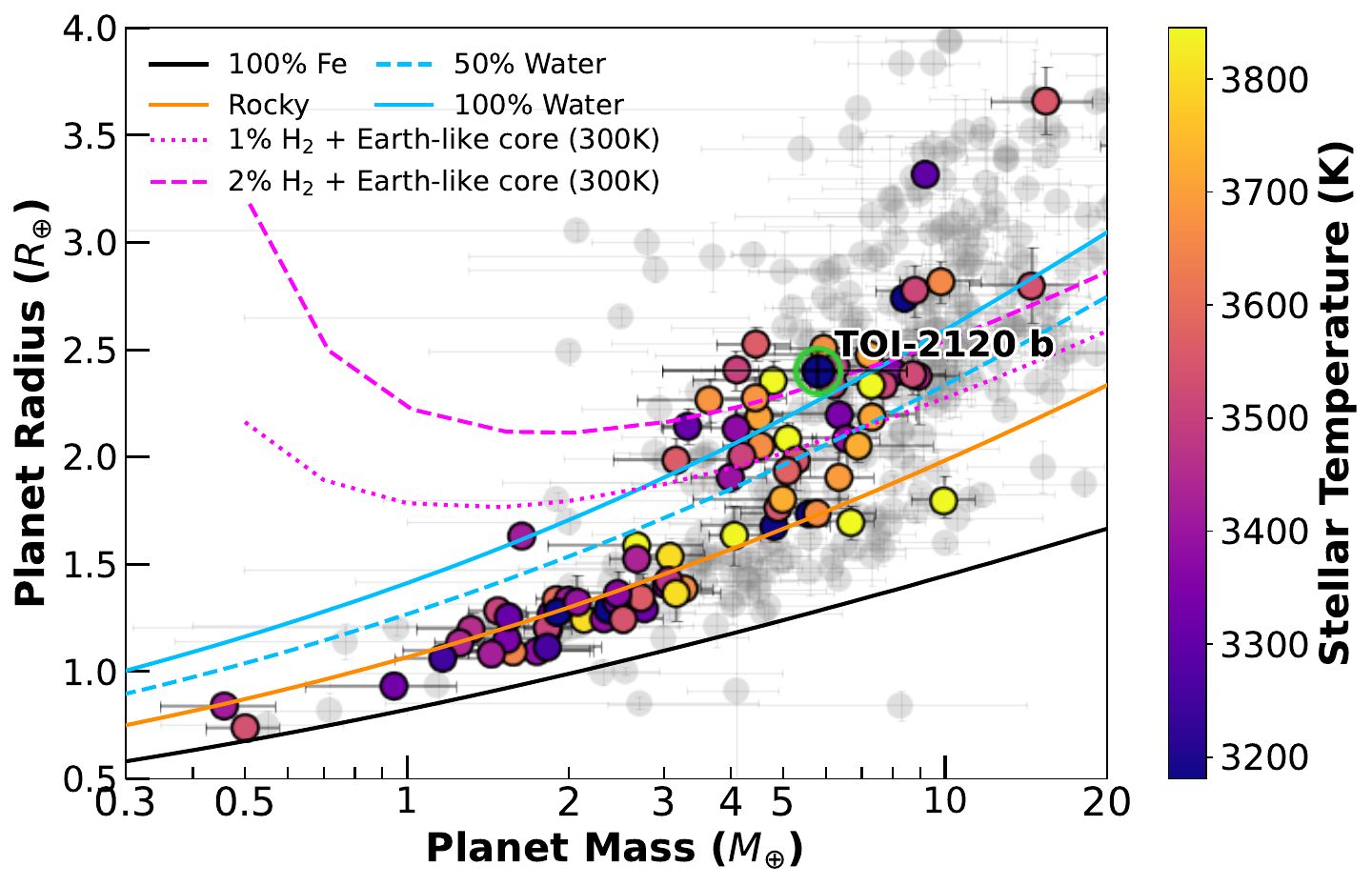}
    \caption{Planet radius vs. mass for TOI-2120\,b (highlighted by a green circle) alongside other planets around M dwarfs with radii $<$4\,\Rearth. The stellar effective temperature is color-coded for each planet, and for comparison, planets around FGK-type stars are displayed in gray. Overplotted are isocomposition curves from \citet{zeng2019} for 100\% Fe (black), rocky (orange), 50\% water (dashed blue), and 100\% water (solid blue), along with 1\% and 2\% H$_2$ envelope curves at 300\,K (magenta, dotted and dashed, respectively).}
    \label{fig:iso}
\end{figure*}

% With a mass of $5.81^{+2.94}_{-2.69}$\,\Mearth and radius of $2.306^{+0.075}_{-0.073}$\,\Rearth, TOI-2120\,b has a bulk density of $2.62^{+1.37}_{-1.22}$ g cm$^{-3}$. As can be seen in Figure~\ref{fig:iso}, its position in mass-radius space is compared with the population of transiting M dwarf planets and with theoretical isocomposition curves from \citet{zeng2019}. The nominal mass and radius place TOI-2120\,b near the volatile-rich portion of the diagram, overlapping the water-rich population identified by \citet{luquepalle2022}. This population-level interpretation is notable because \citet{luquepalle2022} found that small M dwarf planets separate into statistically distinct rocky and lower-density populations, with the latter interpreted as volatile-rich worlds, rather than simply extending the rocky/sub-Neptune radius bimodality observed around FGK stars. Formation models similarly predict that water-rich planets should be preferentially produced around M dwarfs, where embryos forming beyond the snow line can migrate inward while retaining substantial volatile inventories \citep{venturini2024}. TOI-2120\,b is therefore consistent with the emerging picture in which a significant fraction of the low-density planets around M dwarfs acquired their volatile content during formation rather than solely through the retention of primordial H/He envelopes.

We measure the mass of TOI-2120\b to be $5.81^{+2.94}_{-2.69}$\,\Mearth and its radius to $2.306^{+0.075}_{-0.073}$\,\Rearth. In Figure~\ref{fig:iso}, we compare it to the transiting M-dwarf planet population and isocomposition curves of \citet{zeng2019}. TOI-2120\,b lies near the mode of the population of well-characterized M dwarf sub-Neptunes whose bulk density, $\rho=2.62^{+1.37}_{-1.22}\,$g/cm$^{3}$, places it at a confluence of potential compositions. As it must be volatile-rich, this planet could have a H/He envelope or be part of the statistically distinct water world population identified by \citet{luquepalle2022}. Formation models predict that water-rich planets should be preferentially produced around M dwarfs, where embryos forming beyond the snow line can migrate inward while retaining substantial volatile inventories \citep{venturini2024}. TOI-2120\,b is therefore consistent with the emerging picture in which a significant fraction of the low-density planets around M dwarfs acquired their volatile content during formation rather than solely through the retention of primordial H/He envelopes.

Differentiating these two scenarios for this planet is not possible with the present mass uncertainties, however. The importance of the mass precision is underscored by population-level analyses showing that the separation between rocky and volatile-rich M dwarf planets becomes substantially harder to resolve as individual mass uncertainties increase \citep{schulze2024}. A more precise RV mass for TOI-2120\,b could therefore determine whether it occupies the volatile-rich branch of the M dwarf mass-radius distribution or instead remains consistent with a lower volatile inventory.

% The present mass uncertainty, however, prevents a unique compositional interpretation. In particular, the location of TOI-2120\,b relative to the water-rich isocomposition curves should not be interpreted as evidence for a purely water-dominated interior, since comparable radii can be produced by different combinations of a rocky core, volatile-rich material, and a residual H/He envelope. The importance of the mass precision is underscored by population-level analyses showing that the separation between rocky and volatile-rich M dwarf planets becomes substantially harder to resolve as individual mass uncertainties increase \citep{schulze2024}. A more precise RV mass for TOI-2120\,b could therefore determine whether it occupies the volatile-rich branch of the M dwarf mass-radius distribution or instead remains consistent with a lower volatile inventory.

TOI-2120\,b is also an unusually favorable target for testing this interpretation through atmospheric characterization. Its small host star produces a large planet-to-star area ratio and, consequently, a large expected transmission signal. The planet has a Transmission Spectroscopy Metric (TSM; \citealt{kempton2018}) of $\sim$130, well above the threshold of 90 recommended by \citet{kempton2018} for planets of its size. Atmospheric characterization could therefore provide an independent constraint on the planet's volatile inventory.

Several additional observations would further constrain the properties and architecture of this system. First, additional RV measurements would improve the precision of the planet mass, currently measured at 2$\sigma$. A more precise mass would provide a tighter constraint on the bulk density and help distinguish between possible interior compositions. It would also improve the interpretation of a transmission spectrum, as atmospheric retrievals depend on the assumed surface gravity \citep{Batalha2019}. Second, the significant orbital eccentricity ($e \approx 0.32$) is unusual for a planet on a 5.8 day orbit, given the short tidal circularization timescales expected at such orbital separations, and may indicate the presence of a perturbing companion. Continued RV monitoring, together with observations from additional TESS sectors and searches for transit timing variations, could reveal additional planets and test this possibility. Third, transmission spectroscopy with JWST could distinguish between a high-mean-molecular-weight, volatile-rich atmosphere and an extended H/He envelope. Observations of atmospheric escape tracers, such as the metastable He I 10830 \AA triplet, could provide a complementary constraint on the presence of a primordial envelope. These observations will need to account for contamination from unocculted stellar heterogeneities, which can significantly affect transmission spectra of planets orbiting mid-M dwarfs.

Together, these observations could establish whether TOI-2120\,b is representative of the volatile-rich population emerging around the lowest-mass stars or instead belongs to the comparatively rare population of H/He-dominated sub-Neptunes that persist in this regime.

% Together, improved RV constraints and transmission spectroscopy could establish whether TOI-2120\,b is representative of the volatile-rich population emerging around the lowest-mass stars or instead represents one of the comparatively rare H/He-dominated sub-Neptunes that persist in this regime.

\section{Summary}\label{sec:summary}
We present updated mass and radius measurements of TOI-2120\,b, a short-period sub-Neptune orbiting a $0.211\pm0.020$\,\Msun mid-M dwarf. We combine archival TESS and ground-based transit photometry with two additional transit observations obtained with ARCTIC and Red Buttes Observatory, together with new high-precision RV measurements from the Habitable-zone Planet Finder (HPF). We use these observations to jointly constrain the planetary transit and RV signals and investigate the potential contribution of stellar activity to the observed RV variations. We find no significant correlations between the HPF RVs and the activity indicators considered, and the absence of detectable rotational broadening provides a constraint of $v\sin i<2$\,km s$^{-1}$. Together, these results suggest that stellar activity is unlikely to be the dominant source of variability in the HPF RV measurements.

Our joint analysis yields a planetary radius of $2.306^{+0.075}_{-0.073}$\,\Rearth and a mass of $5.81^{+2.94}_{-2.69}$\,\Mearth, corresponding to a bulk density of $2.62^{+1.37}_{-1.22}$ g cm$^{-3}$. The revised radius is larger than the value reported by \citet{Hori2024}, although the two measurements remain consistent within $2\sigma$, and we attribute the difference primarily to the additional ARCTIC and RBO transit observations. Our HPF RV measurements provide an independent, quantitative constraint on the planetary mass of $5.81^{+2.94}_{-2.69}$\,\Mearth, consistent with the $2\sigma$ upper limit of $<$6.8\,\Mearth reported by \citet{Hori2024} from Subaru/IRD observations. The resulting low bulk density is consistent with a water-rich composition, although the relatively large mass uncertainty prevents us from making detailed compositional inferences. We also find a modest eccentricity of $e=0.140^{+0.129}_{-0.087}$, consistent within $2\sigma$ with the higher eccentricity reported by \citet{Hori2024}. However, our fit provides only a $1.6\sigma$ preference for a non-zero eccentricity, and therefore the current data do not distinguish between a modestly eccentric and circular orbit.

TOI-2120\,b is among the small number of sub-Neptunes orbiting very cool stars for which both mass and radius have been measured, making it a useful system for investigating the emerging population of small, low-density planets around M dwarfs. The combination of its sub-Neptune radius and small host star therefore makes TOI-2120\,b a promising target for atmospheric characterization with James Webb Space Telescope (JWST). Improving the mass precision of TOI-2120\,b would strengthen both the interior and atmospheric interpretations of the system. If its water-rich bulk composition is confirmed, transmission spectroscopy could provide an independent constraint on its volatile inventory and formation history, linking the atmospheric properties of TOI-2120\,b to the broader population of volatile-rich planets emerging around low-mass stars. Additional RV observations will therefore be important for further constraining its mass and composition.

% The combination of its sub-Neptune radius and small host star therefore makes TOI-2120,b a promising target for atmospheric characterization with JWST. Improving the mass precision of TOI-2120,b would therefore strengthen both the interior and atmospheric interpretations of the system. If its water-rich bulk composition is confirmed, transmission spectroscopy could provide an independent constraint on its volatile inventory and formation history, linking the atmospheric properties of TOI-2120,b to the broader population of volatile-rich planets emerging around low-mass stars.

%%%%%%%%%%%%%%%%%%%%%%%%%%%%%%%%%%%%%%%%%%%%%%%%%%%%%%%
%%%%%%%%%%%%%%%%%%% Facilities %%%%%%%%%%%%%%%%%%%%%%%
%%%%%%%%%%%%%%%%%%%%%%%%%%%%%%%%%%%%%%%%%%%%%%%%%%%%%%%
\facilities{
HET (HPF), ARC (ARCTIC), Red Buttes 0.6m, Shane (ShaneAO), TESS, MAST, Gaia, NASA Exoplanet Archive}

%%%%%%%%%%%%%%%%%%%%%%%%%%%%%%%%%%%%%%%%%%%%%%%%%%%%%%%
%%%%%%%%%%%%%%%%%%% ACKNOWLEDGMENTS %%%%%%%%%%%%%%%%%%%%%%%
%%%%%%%%%%%%%%%%%%%%%%%%%%%%%%%%%%%%%%%%%%%%%%%%%%%%%%%

\section{Acknowledgments}
R.B.F. would like to thank Suvrath Mahadevan, Leslie Hebb, and Adam Kowalski for their expertise, insightful discussions, and helpful comments that contributed to this work.
These findings stem from observations conducted using the Habitable-zone Planet Finder Spectrograph on the Hobby-Eberly Telescope (HET). Furthermore, we acknowledge the collaborative efforts of the University of Texas at Austin, the Pennsylvania State University, Ludwig-Maximilian-Universität München, and Georg-August Universität Gottingen in the Hobby–Eberly Telescope project. The HET is named in recognition of its principal benefactors, William P. Hobby and Robert E. Eberly.
The HET collaboration extends its appreciation to the Texas Advanced Computing Center for its support and resources. We express gratitude to the Resident Astronomers and Telescope Operators at the HET for their adept execution of observations with HPF. Additionally, we acknowledge that the HET is situated on Indigenous land. Furthermore, we wish to recognize and pay our respects to the Carrizo \& Comecrudo, Coahuiltecan, Caddo, Tonkawa, Comanche, Lipan Apache, Alabama-Coushatta, Kickapoo, Tigua Pueblo, and all the American Indian and Indigenous Peoples and communities who have inhabited or become part of these lands and territories in Texas, here on Turtle Island. We acknowledge the Texas Advanced Computing Center (TACC) at The University of Texas at Austin for providing high-performance computing, visualization, and storage resources that have contributed to the results reported within this paper.

This work was partially supported by funding from the Center for Exoplanets and Habitable Worlds. The Center for Exoplanets and Habitable Worlds and the Penn State Extraterrestrial Intelligence Center are supported by Penn State and its Eberly College of Science.  Computations for this research were performed on the Penn State’s Institute for Computational and Data Sciences' Advanced CyberInfrastructure (ICDS-ACI). This content is solely the responsibility of the authors and does not necessarily represent the views of the Institute for Computational and Data Sciences.

We utilized data from the Gaia mission\footnote{\url{https://www.cosmos.esa.int/gaia}} of European Space Agency (ESA), processed by the Gaia Data Processing and Analysis Consortium (DPAC, \footnote{\url{https://www.cosmos.esa.int/web/gaia/dpac/consortium}}). The DPAC's funding is provided by national institutions, particularly those involved in the Gaia Multilateral Agreement. Additionally, we acknowledge support from NSF grant AST-1907622 for conducting precise photometric observations from the ground.  P.~R. was supported by NSF grant AST-1910954 in conducting ShaneAO observations.

CIC acknowledges support by (i) NASA Headquarters through award number 80GSFC24M0006 and (ii) the GSFC Sellers Exoplanet Environments Collaboration (SEEC), which is supported by NASA's Planetary, Astrophysics and Heliophysics Science Divisions' Research Program. S.P.H acknowledges that this research was carried out, in part, at the Jet Propulsion Laboratory, California Institute of Technology, under a contract with the National Aeronautics and Space Administration (80NM0018D0004).

\clearpage
%%%%%%%%%%%%%%%%%%%%%%%%%%%%%%%%%%%%%%%%%%%%%%%%%%%%%%%
%%%%%%%%%%%%%%%%%%% BIBLIOGRAPHY %%%%%%%%%%%%%%%%%%%%%%%
%%%%%%%%%%%%%%%%%%%%%%%%%%%%%%%%%%%%%%%%%%%%%%%%%%%%%%%

\bibliographystyle{aasjournalv7.1.bst}
\bibliography{main}

@misc{PSCompPars,
  author       = {{NASA Exoplanet Archive}},
  title        = {Planetary Systems Composite Parameters},
  publisher    = {NExScI-Caltech/IPAC},
  year         = {2026},
  month        = oct,
  note         = {Version: 2026-10-DD},
  doi          = {10.26133/NEA13},
  url          = {https://doi.org/10.26133/NEA13}
}

@ARTICLE{Christiansen2025,
       author = {{Christiansen}, Jessie L. and {McElroy}, Douglas L. and {Harbut}, Marcy and {Ciardi}, David R. and {Crane}, Megan and {Good}, John and {Hardegree-Ullman}, Kevin K. and {Kesseli}, Aurora Y. and {Lund}, Michael B. and {Lynn}, Meca and et al.},
        title = "{The NASA Exoplanet Archive and Exoplanet Follow-up Observing Program: Data, Tools, and Usage}",
      journal = {\psj},
         year = 2025,
        month = aug,
       volume = {6},
       number = {8},
          eid = {186},
        pages = {186},
          doi = {10.3847/PSJ/ade3c2},
archivePrefix = {arXiv},
       eprint = {2506.03299},
 primaryClass = {astro-ph.EP},
       adsurl = {https://scixplorer.org/abs/2025PSJ.....6..186C}
}

@ARTICLE{lubin2021,
       author = {{Lubin}, Jack and {Robertson}, Paul and {Stefansson}, Gudmundur and {Ninan}, Joe and {Mahadevan}, Suvrath and {Endl}, Michael and {Ford}, Eric and {Wright}, Jason T. and {Beard}, Corey and {Bender}, Chad and {Cochran}, William D. and {Diddams}, Scott A. and {Fredrick}, Connor and {Halverson}, Samuel and {Kanodia}, Shubham and {Metcalf}, Andrew J. and {Ramsey}, Lawrence and {Roy}, Arpita and {Schwab}, Christian and {Terrien}, Ryan},
        title = "{Stellar Activity Manifesting at a One-year Alias Explains Barnard b as a False Positive}",
      journal = {\aj},
         year = 2021,
        month = aug,
       volume = {162},
       number = {2},
          eid = {61},
        pages = {61},
          doi = {10.3847/1538-3881/ac0057},
       adsurl = {https://ui.adsabs.harvard.edu/abs/2021AJ....162...61L}
}

@article{kendall1948rank,
  title={Rank correlation methods.},
  author={Kendall, Maurice George},
  year={1948},
  publisher={Griffin}
}

@ARTICLE{RicheyYowell2023,
       author = {{Richey-Yowell}, Tyler and {Shkolnik}, Evgenya L. and {Schneider}, Adam C. and {Peacock}, Sarah and {Huseby}, Lori A. and {Jackman}, James A.~G. and {Barman}, Travis and {Osby}, Ella and {Meadows}, Victoria S.},
        title = "{HAZMAT. IX. An Analysis of the UV and X-Ray Evolution of Low-mass Stars in the Era of Gaia}",
      journal = {\apj},
         year = 2023,
        month = jul,
       volume = {951},
       number = {1},
          eid = {44},
        pages = {44},
          doi = {10.3847/1538-4357/acd2dc},
archivePrefix = {arXiv},
       eprint = {2305.06561},
 primaryClass = {astro-ph.SR},
       adsurl = {https://ui.adsabs.harvard.edu/abs/2023ApJ...951...44R}
}

@ARTICLE{Peacock2020,
       author = {{Peacock}, Sarah and {Barman}, Travis and {Shkolnik}, Evgenya L. and {Loyd}, R.~O. Parke and {Schneider}, Adam C. and {Pagano}, Isabella and {Meadows}, Victoria S.},
        title = "{HAZMAT VI: The Evolution of Extreme Ultraviolet Radiation Emitted from Early M Stars}",
      journal = {\apj},
         year = 2020,
        month = may,
       volume = {895},
       number = {1},
          eid = {5},
        pages = {5},
          doi = {10.3847/1538-4357/ab893a},
archivePrefix = {arXiv},
       eprint = {2005.01687},
 primaryClass = {astro-ph.SR},
       adsurl = {https://ui.adsabs.harvard.edu/abs/2020ApJ...895....5P}
}

@ARTICLE{Gaiadr3,
       author = {{Fouesneau}, M. and {Fr{\'e}mat}, Y. and {Andrae}, R. and {Korn}, A.~J. and {Soubiran}, C. and {Kordopatis}, G. and {Vallenari}, A. and {Heiter}, U. and {Creevey}, O.~L. and {Sarro}, L.~M. and {de Laverny}, P. and {Lanzafame}, A.~C. and {Lobel}, A. and {Sordo}, R. and {Rybizki}, J. and {Slezak}, I. and {{\'A}lvarez}, M.~A. and {Drimmel}, R. and {Garabato}, D. and {Delchambre}, L. and {Bailer-Jones}, C.~A.~L. and {Hatzidimitriou}, D. and {Lorca}, A. and {Le Fustec}, Y. and {Pailler}, F. and {Mary}, N. and {Robin}, C. and {Utrilla}, E. and {Abreu Aramburu}, A. and {Bakker}, J. and {Bellas-Velidis}, I. and {Bijaoui}, A. and {Blomme}, R. and {Bouret}, J.-C. and {Brouillet}, N. and {Brugaletta}, E. and {Burlacu}, A. and {Carballo}, R. and {Casamiquela}, L. and {Chaoul}, L. and {Chiavassa}, A. and {Contursi}, G. and {Cooper}, W.~J. and {Dafonte}, C. and {Demouchy}, C. and {Dharmawardena}, T.~E. and {Garc{\'\i}a-Lario}, P. and {Garc{\'\i}a-Torres}, M. and {Gomez}, A. and {Gonz{\'a}lez-Santamar{\'\i}a}, I. and {Jean-Antoine Piccolo}, A. and {Kontizas}, M. and {Lebreton}, Y. and {Licata}, E.~L. and {Lindstr{\o}m}, H.~E.~P. and {Livanou}, E. and {Magdaleno Romeo}, A. and {Manteiga}, M. and {Marocco}, F. and {Martayan}, C. and {Marshall}, D.~J. and {Nicolas}, C. and {Ordenovic}, C. and {Palicio}, P.~A. and {Pallas-Quintela}, L. and {Pichon}, B. and {Poggio}, E. and {Recio-Blanco}, A. and {Riclet}, F. and {Santove{\~n}a}, R. and {Schultheis}, M.~S. and {Segol}, M. and {Silvelo}, A. and {Smart}, R.~L. and {S{\"u}veges}, M. and {Th{\'e}venin}, F. and {Torralba Elipe}, G. and {Ulla}, A. and {van Dillen}, E. and {Zhao}, H. and {Zorec}, J.},
        title = "{Gaia Data Release 3. Apsis. II. Stellar parameters}",
      journal = {\aap},
         year = 2023,
        month = jun,
       volume = {674},
          eid = {A28},
        pages = {A28},
          doi = {10.1051/0004-6361/202243919},
archivePrefix = {arXiv},
       eprint = {2206.05992},
 primaryClass = {astro-ph.SR},
       adsurl = {https://ui.adsabs.harvard.edu/abs/2023A&A...674A..28F}
}

@ARTICLE{kempton2018,
       author = {{Kempton}, Eliza M.-R. and {Bean}, Jacob L. and {Louie}, Dana R. and {Deming}, Drake and {Koll}, Daniel D.~B. and {Mansfield}, Megan and {Christiansen}, Jessie L. and {L{\'o}pez-Morales}, Mercedes and {Swain}, Mark R. and {Zellem}, Robert T. and {Ballard}, Sarah and {Barclay}, Thomas and {Barstow}, Joanna K. and {Batalha}, Natasha E. and {Beatty}, Thomas G. and {Berta-Thompson}, Zach and {Birkby}, Jayne and {Buchhave}, Lars A. and {Charbonneau}, David and {Cowan}, Nicolas B. and {Crossfield}, Ian and {de Val-Borro}, Miguel and {Doyon}, Ren{\'e} and {Dragomir}, Diana and {Gaidos}, Eric and {Heng}, Kevin and {Hu}, Renyu and {Kane}, Stephen R. and {Kreidberg}, Laura and {Mallonn}, Matthias and {Morley}, Caroline V. and {Narita}, Norio and {Nascimbeni}, Valerio and {Pall{\'e}}, Enric and {Quintana}, Elisa V. and {Rauscher}, Emily and {Seager}, Sara and {Shkolnik}, Evgenya L. and {Sing}, David K. and {Sozzetti}, Alessandro and {Stassun}, Keivan G. and {Valenti}, Jeff A. and {von Essen}, Carolina},
        title = "{A Framework for Prioritizing the TESS Planetary Candidates Most Amenable to Atmospheric Characterization}",
      journal = {\pasp},
         year = 2018,
        month = nov,
       volume = {130},
       number = {993},
        pages = {114401},
          doi = {10.1088/1538-3873/aadf6f},
archivePrefix = {arXiv},
       eprint = {1805.03671},
 primaryClass = {astro-ph.EP},
       adsurl = {https://ui.adsabs.harvard.edu/abs/2018PASP..130k4401K}
}

@ARTICLE{schulze2024,
       author = {{Schulze}, J.~G. and {Wang}, Ji and {Johnson}, J.~A. and {Gaudi}, B.~S. and {Rodriguez Martinez}, R. and {Unterborn}, C.~T. and {Panero}, W.~R.},
        title = "{A Gap in the Densities of Small Planets Orbiting M Dwarfs: Rigorous Statistical Confirmation Using the Open-source Code RhoPop}",
      journal = {\psj},
         year = 2024,
        month = mar,
       volume = {5},
       number = {3},
          eid = {71},
        pages = {71},
          doi = {10.3847/PSJ/ad26f5},
archivePrefix = {arXiv},
       eprint = {2403.13961},
 primaryClass = {astro-ph.EP},
       adsurl = {https://ui.adsabs.harvard.edu/abs/2024PSJ.....5...71S}
}

@ARTICLE{zeng2019,
       author = {{Zeng}, Li and {Jacobsen}, Stein B. and {Sasselov}, Dimitar D. and {Petaev}, Michail I. and {Vanderburg}, Andrew and {Lopez-Morales}, Mercedes and {Perez-Mercader}, Juan and {Mattsson}, Thomas R. and {Li}, Gongjie and {Heising}, Matthew Z. and {Bonomo}, Aldo S. and {Damasso}, Mario and {Berger}, Travis A. and {Cao}, Hao and {Levi}, Amit and {Wordsworth}, Robin D.},
        title = "{Growth model interpretation of planet size distribution}",
      journal = {Proceedings of the National Academy of Science},
         year = 2019,
        month = may,
       volume = {116},
       number = {20},
        pages = {9723-9728},
          doi = {10.1073/pnas.1812905116},
archivePrefix = {arXiv},
       eprint = {1906.04253},
 primaryClass = {astro-ph.EP},
       adsurl = {https://ui.adsabs.harvard.edu/abs/2019PNAS..116.9723Z}
}

@ARTICLE{vaneylen2021,
       author = {{Van Eylen}, V. and {Astudillo-Defru}, N. and {Bonfils}, X. and {Livingston}, J. and {Hirano}, T. and {Luque}, R. and {Lam}, K.~W.~F. and {Justesen}, A.~B. and {Winn}, J.~N. and {Gandolfi}, D. and {Nowak}, G. and {Palle}, E. and {Albrecht}, S. and {Dai}, F. and {Campos Estrada}, B. and {Owen}, J.~E. and {Foreman-Mackey}, D. and {Fridlund}, M. and {Korth}, J. and {Mathur}, S. and {Forveille}, T. and {Mikal-Evans}, T. and {Osborne}, H.~L.~M. and {Ho}, C.~S.~K. and {Almenara}, J.~M. and {Artigau}, E. and {Barrag{\'a}n}, O. and {Barros}, S.~C.~C. and {Bouchy}, F. and {Cabrera}, J. and {Caldwell}, D.~A. and {Charbonneau}, D. and {Chaturvedi}, P. and {Cochran}, W.~D. and {Csizmadia}, S. and {Damasso}, M. and {Delfosse}, X. and {De Medeiros}, J.~R. and {D{\'\i}az}, R.~F. and {Doyon}, R. and {Esposito}, M. and {F{\H{u}}r{\'e}sz}, G. and {Figueira}, P. and {Georgieva}, I. and {Goffo}, E. and {Grziwa}, S. and {Guenther}, E. and {Hatzes}, A.~P. and {Jenkins}, J.~M. and {Kabath}, P. and {Knudstrup}, E. and {Latham}, D.~W. and {Lavie}, B. and {Lovis}, C. and {Mennickent}, R.~E. and {Mullally}, S.~E. and {Murgas}, F. and {Narita}, N. and {Pepe}, F.~A. and {Persson}, C.~M. and {Redfield}, S. and {Ricker}, G.~R. and {Santos}, N.~C. and {Seager}, S. and {Serrano}, L.~M. and {Smith}, A.~M.~S. and {Su{\'a}rez Mascare{\~n}o}, A. and {Subjak}, J. and {Twicken}, J.~D. and {Udry}, S. and {Vanderspek}, R. and {Zapatero Osorio}, M.~R.},
        title = "{Masses and compositions of three small planets orbiting the nearby M dwarf L231-32 (TOI-270) and the M dwarf radius valley}",
      journal = {\mnras},
         year = 2021,
        month = oct,
       volume = {507},
       number = {2},
        pages = {2154-2173},
          doi = {10.1093/mnras/stab2143},
archivePrefix = {arXiv},
       eprint = {2101.01593},
 primaryClass = {astro-ph.EP},
       adsurl = {https://ui.adsabs.harvard.edu/abs/2021MNRAS.507.2154V}
}

@ARTICLE{venturini2024,
       author = {{Venturini}, J. and {Ronco}, M.~P. and {Guilera}, O.~M. and {Haldemann}, J. and {Mordasini}, C. and {Miller Bertolami}, M.},
        title = "{A fading radius valley towards M dwarfs, a persistent density valley across stellar types}",
      journal = {\aap},
         year = 2024,
        month = jun,
       volume = {686},
          eid = {L9},
        pages = {L9},
          doi = {10.1051/0004-6361/202349088},
archivePrefix = {arXiv},
       eprint = {2404.01967},
 primaryClass = {astro-ph.EP},
       adsurl = {https://ui.adsabs.harvard.edu/abs/2024A&A...686L...9V}
}

@ARTICLE{luquepalle2022,
       author = {{Luque}, Rafael and {Pall{\'e}}, Enric},
        title = "{Density, not radius, separates rocky and water-rich small planets orbiting M dwarf stars}",
      journal = {Science},
         year = 2022,
        month = sep,
       volume = {377},
       number = {6611},
        pages = {1211-1214},
          doi = {10.1126/science.abl7164},
archivePrefix = {arXiv},
       eprint = {2209.03871},
 primaryClass = {astro-ph.EP},
       adsurl = {https://ui.adsabs.harvard.edu/abs/2022Sci...377.1211L}
}

@ARTICLE{alibert2013,
       author = {{Alibert}, Y. and {Carron}, F. and {Fortier}, A. and {Pfyffer}, S. and {Benz}, W. and {Mordasini}, C. and {Swoboda}, D.},
        title = "{Theoretical models of planetary system formation: mass vs. semi-major axis}",
      journal = {\aap},
         year = 2013,
        month = oct,
       volume = {558},
          eid = {A109},
        pages = {A109},
          doi = {10.1051/0004-6361/201321690},
archivePrefix = {arXiv},
       eprint = {1307.4864},
 primaryClass = {astro-ph.EP},
       adsurl = {https://ui.adsabs.harvard.edu/abs/2013A&A...558A.109A}
}

@ARTICLE{Venturini2020,
       author = {{Venturini}, Julia and {Guilera}, Octavio M. and {Haldemann}, Jonas and {Ronco}, Mar{\'\i}a P. and {Mordasini}, Christoph},
        title = "{The nature of the radius valley. Hints from formation and evolution models}",
      journal = {\aap},
         year = 2020,
        month = nov,
       volume = {643},
          eid = {L1},
        pages = {L1},
          doi = {10.1051/0004-6361/202039141},
archivePrefix = {arXiv},
       eprint = {2008.05513},
 primaryClass = {astro-ph.EP},
       adsurl = {https://ui.adsabs.harvard.edu/abs/2020A&A...643L...1V}
}

@ARTICLE{Bitsch2016,
       author = {{Bitsch}, Bertram and {Johansen}, Anders},
        title = "{Influence of the water content in protoplanetary discs on planet migration and formation}",
      journal = {\aap},
         year = 2016,
        month = may,
       volume = {590},
          eid = {A101},
        pages = {A101},
          doi = {10.1051/0004-6361/201527676},
archivePrefix = {arXiv},
       eprint = {1603.01125},
 primaryClass = {astro-ph.EP},
       adsurl = {https://ui.adsabs.harvard.edu/abs/2016A&A...590A.101B}
}

@ARTICLE{raymond2018water,
       author = {{Raymond}, Sean N. and {Boulet}, Thibault and {Izidoro}, Andre and
         {Esteves}, Leandro and {Bitsch}, Bertram},
        title = "{Migration-driven diversity of super-Earth compositions}",
      journal = {\mnras},
         year = 2018,
        month = sep,
       volume = {479},
       number = {1},
        pages = {L81-L85},
          doi = {10.1093/mnrasl/sly100},
archivePrefix = {arXiv},
       eprint = {1805.10345},
 primaryClass = {astro-ph.EP},
       adsurl = {https://ui.adsabs.harvard.edu/abs/2018MNRAS.479L..81R}
}

@article{gupta2019sculpting,
  title={Sculpting the valley in the radius distribution of small exoplanets as a by-product of planet formation: the core-powered mass-loss mechanism},
  author={Gupta, Akash and Schlichting, Hilke E},
  journal={\mnras},
  volume={487},
  number={1},
  pages={24--33},
  year={2019},
  publisher={Oxford University Press}
}

@article{ginzburg2018core,
  title={Core-powered mass-loss and the radius distribution of small exoplanets},
  author={Ginzburg, Sivan and Schlichting, Hilke E and Sari, Re'em},
  journal={\mnras},
  volume={476},
  number={1},
  pages={759--765},
  year={2018},
  publisher={Oxford University Press}
}

@ARTICLE{owenwu2017,
       author = {{Owen}, James E. and {Wu}, Yanqin},
        title = "{The Evaporation Valley in the Kepler Planets}",
      journal = {\apj},
         year = 2017,
        month = sep,
       volume = {847},
       number = {1},
          eid = {29},
        pages = {29},
          doi = {10.3847/1538-4357/aa890a},
archivePrefix = {arXiv},
       eprint = {1705.10810},
 primaryClass = {astro-ph.EP},
       adsurl = {https://ui.adsabs.harvard.edu/abs/2017ApJ...847...29O}
}

@ARTICLE{gillis2026,
       author = {{Gillis}, Erik Diego and {Cloutier}, Ryan and {Pass}, Emily K.},
        title = "{TESS Planet Occurrence Rates Reveal the Disappearance of the Radius Valley around Mid-to-late M Dwarfs}",
      journal = {\aj},
         year = 2026,
        month = may,
       volume = {171},
       number = {5},
          eid = {317},
        pages = {317},
          doi = {10.3847/1538-3881/ae5810},
archivePrefix = {arXiv},
       eprint = {2602.23364},
 primaryClass = {astro-ph.EP},
       adsurl = {https://ui.adsabs.harvard.edu/abs/2026AJ....171..317G}
}

@ARTICLE{cloutiermenou2020,
       author = {{Cloutier}, Ryan and {Menou}, Kristen},
        title = "{Evolution of the Radius Valley around Low-mass Stars from Kepler and K2}",
      journal = {\aj},
         year = 2020,
        month = may,
       volume = {159},
       number = {5},
          eid = {211},
        pages = {211},
          doi = {10.3847/1538-3881/ab8237},
archivePrefix = {arXiv},
       eprint = {1912.02170},
 primaryClass = {astro-ph.EP},
       adsurl = {https://ui.adsabs.harvard.edu/abs/2020AJ....159..211C}
}

@ARTICLE{berger2020,
       author = {{Berger}, Travis A. and {Huber}, Daniel and {Gaidos}, Eric and {van Saders}, Jennifer L. and {Weiss}, Lauren M.},
        title = "{The Gaia-Kepler Stellar Properties Catalog. II. Planet Radius Demographics as a Function of Stellar Mass and Age}",
      journal = {\aj},
         year = 2020,
        month = sep,
       volume = {160},
       number = {3},
          eid = {108},
        pages = {108},
          doi = {10.3847/1538-3881/aba18a},
archivePrefix = {arXiv},
       eprint = {2005.14671},
 primaryClass = {astro-ph.EP},
       adsurl = {https://ui.adsabs.harvard.edu/abs/2020AJ....160..108B}
}

@ARTICLE{Petigura2022,
       author = {{Petigura}, Erik A. and {Rogers}, James G. and {Isaacson}, Howard and {Owen}, James E. and {Kraus}, Adam L. and {Winn}, Joshua N. and {MacDougall}, Mason G. and {Howard}, Andrew W. and {Fulton}, Benjamin and {Kosiarek}, Molly R. and {Weiss}, Lauren M. and {Behmard}, Aida and {Blunt}, Sarah},
        title = "{The California-Kepler Survey. X. The Radius Gap as a Function of Stellar Mass, Metallicity, and Age}",
      journal = {\aj},
         year = 2022,
        month = apr,
       volume = {163},
       number = {4},
          eid = {179},
        pages = {179},
          doi = {10.3847/1538-3881/ac51e3},
archivePrefix = {arXiv},
       eprint = {2201.10020},
 primaryClass = {astro-ph.EP},
       adsurl = {https://ui.adsabs.harvard.edu/abs/2022AJ....163..179P}
}

@article{hardegree2020scaling,
  title={Scaling K2. I. Revised parameters for 222,088 K2 stars and a K2 planet radius valley at 1.9 R⊕},
  author={Hardegree-Ullman, Kevin K and Zink, Jon K and Christiansen, Jessie L and Dressing, Courtney D and Ciardi, David R and Schlieder, Joshua E},
  journal={\apjs},
  volume={247},
  number={1},
  pages={28},
  year={2020},
  publisher={IOP Publishing}
}

@article{fulton2017california,
  title={The California-Kepler survey. III. A gap in the radius distribution of small planets},
  author={Fulton, Benjamin J and Petigura, Erik A and Howard, Andrew W and Isaacson, Howard and Marcy, Geoffrey W and Cargile, Phillip A and Hebb, Leslie and Weiss, Lauren M and Johnson, John Asher and Morton, Timothy D and others},
  journal={\aj},
  volume={154},
  number={3},
  pages={109},
  year={2017},
  publisher={IOP Publishing}
}

@ARTICLE{Borucki2010,
       author = {{Borucki}, William J. and {Koch}, David and {Basri}, Gibor and
         {Batalha}, Natalie and {Brown}, Timothy and {Caldwell}, Douglas and
         {Caldwell}, John and {Christensen-Dalsgaard}, J{\o}rgen and
         {Cochran}, William D. and {DeVore}, Edna and {Dunham}, Edward W. and
         {Dupree}, Andrea K. and {Gautier}, Thomas N. and {Geary}, John C. and
         {Gilliland}, Ronald and {Gould}, Alan and {Howell}, Steve B. and
         {Jenkins}, Jon M. and {Kondo}, Yoji and {Latham}, David W. and
         {Marcy}, Geoffrey W. and {Meibom}, S{\o}ren and {Kjeldsen}, Hans and
         {Lissauer}, Jack J. and {Monet}, David G. and {Morrison}, David and
         {Sasselov}, Dimitar and {Tarter}, Jill and {Boss}, Alan and
         {Brownlee}, Don and {Owen}, Toby and {Buzasi}, Derek and
         {Charbonneau}, David and {Doyle}, Laurance and {Fortney}, Jonathan and
         {Ford}, Eric B. and {Holman}, Matthew J. and {Seager}, Sara and
         {Steffen}, Jason H. and {Welsh}, William F. and {Rowe}, Jason and
         {Anderson}, Howard and {Buchhave}, Lars and {Ciardi}, David and
         {Walkowicz}, Lucianne and {Sherry}, William and {Horch}, Elliott and
         {Isaacson}, Howard and {Everett}, Mark E. and {Fischer}, Debra and
         {Torres}, Guillermo and {Johnson}, John Asher and {Endl}, Michael and
         {MacQueen}, Phillip and {Bryson}, Stephen T. and {Dotson}, Jessie and
         {Haas}, Michael and {Kolodziejczak}, Jeffrey and {Van Cleve}, Jeffrey and
         {Chandrasekaran}, Hema and {Twicken}, Joseph D. and
         {Quintana}, Elisa V. and {Clarke}, Bruce D. and {Allen}, Christopher and
         {Li}, Jie and {Wu}, Haley and {Tenenbaum}, Peter and
         {Verner}, Ekaterina and {Bruhweiler}, Frederick and {Barnes}, Jason and
         {Prsa}, Andrej},
        title = "{Kepler Planet-Detection Mission: Introduction and First Results}",
      journal = {Science},
         year = 2010,
        month = feb,
       volume = {327},
       number = {5968},
        pages = {977},
          doi = {10.1126/science.1185402},
       adsurl = {https://ui.adsabs.harvard.edu/abs/2010Sci...327..977B}
}

@ARTICLE{Gore2024,
       author = {{Gore}, Rebecca and {Giacalone}, Steven and {Dressing}, Courtney D. and {Turtelboom}, Emma V. and {Schroeder}, Ashley and {Fortenbach}, Charles D. and {Hardegree-Ullman}, Kevin K. and {Zink}, Jon K. and {Mayo}, Andrew W. and {Schlieder}, Joshua E. and {Christiansen}, Jessie L.},
        title = "{Metallicities and Refined Stellar Parameters for 52 Cool Dwarfs with Transiting Planets and Planet Candidates}",
      journal = {\apjs},
         year = 2024,
        month = apr,
       volume = {271},
       number = {2},
          eid = {48},
        pages = {48},
          doi = {10.3847/1538-4365/ad2c0c},
archivePrefix = {arXiv},
       eprint = {2402.13223},
 primaryClass = {astro-ph.EP},
       adsurl = {https://ui.adsabs.harvard.edu/abs/2024ApJS..271...48G}
}

@ARTICLE{Hirano2020,
       author = {{Hirano}, Teruyuki and {Kuzuhara}, Masayuki and {Kotani}, Takayuki and {Omiya}, Masashi and {Kudo}, Tomoyuki and {Harakawa}, Hiroki and {Vievard}, S{\'e}bastien and {Kurokawa}, Takashi and {Nishikawa}, Jun and {Tamura}, Motohide and {Hodapp}, Klaus and {Ishizuka}, Masato and {Jacobson}, Shane and {Konishi}, Mihoko and {Serizawa}, Takuma and {Ueda}, Akitoshi and {Gaidos}, Eric and {Sato}, Bun'ei},
        title = "{Precision radial velocity measurements by the forward-modeling technique in the near-infrared}",
      journal = {\pasj},
         year = 2020,
        month = dec,
       volume = {72},
       number = {6},
          eid = {93},
        pages = {93},
          doi = {10.1093/pasj/psaa085},
archivePrefix = {arXiv},
       eprint = {2007.11013},
 primaryClass = {astro-ph.EP},
       adsurl = {https://ui.adsabs.harvard.edu/abs/2020PASJ...72...93H}
}

@INPROCEEDINGS{Kuzuhara2018,
       author = {{Kuzuhara}, Masayuki and {Hirano}, Teruyuki and {Kotani}, Takayuki and {Ishizuka}, Masato and {Omiya}, Masashi and {Konishi}, Mihoko and {Kudo}, Tomoyuki and {Nishikawa}, Jun and {Ueda}, Akitoshi and {Hosokawa}, Ko and {Kusakabe}, Nobuhiko and {Kurokawa}, Takashi and {Kokubo}, Tsukasa and {Mori}, Takahiro and {Tanaka}, Yosuke and {Jacobson}, Shane and {Hodapp}, Klaus and {Tamura}, Motohide},
        title = "{Performance tests of Subaru/IRD for very precise and stable infrared radial velocity observations}",
    booktitle = {Ground-based and Airborne Instrumentation for Astronomy VII},
         year = 2018,
       editor = {{Evans}, Christopher J. and {Simard}, Luc and {Takami}, Hideki},
       series = {Society of Photo-Optical Instrumentation Engineers (SPIE) Conference Series},
       volume = {10702},
        month = jul,
          eid = {1070260},
        pages = {1070260},
          doi = {10.1117/12.2311832},
       adsurl = {https://ui.adsabs.harvard.edu/abs/2018SPIE10702E..60K}
}

@INPROCEEDINGS{Kotani2018,
       author = {{Kotani}, Takayuki and {Tamura}, Motohide and {Nishikawa}, Jun and {Ueda}, Akitoshi and {Kuzuhara}, Masayuki and {Omiya}, Masashi and {Hashimoto}, Jun and {Ishizuka}, Masato and {Hirano}, Teruyuki and {Suto}, Hiroshi and {Kurokawa}, Takashi and {Kokubo}, Tsukasa and {Mori}, Takahiro and {Tanaka}, Yosuke and {Kashiwagi}, Ken and {Konishi}, Mihoko and {Kudo}, Tomoyuki and {Sato}, Bun'ei and {Jacobson}, Shane and {Hodapp}, Klaus W. and {Hall}, Donald B. and {Aoki}, Wako and {Usuda}, Tomonori and {Nishiyama}, Shogo and {Nakajima}, Tadashi and {Ikeda}, Yuji and {Yamamuro}, Tomoyasu and {Morino}, Jun-Ichi and {Baba}, Haruka and {Hosokawa}, Ko and {Ishikawa}, Hiroyuki and {Narita}, Norio and {Kokubo}, Eiichiro and {Hayano}, Yutaka and {Izumiura}, Hideyuki and {Kambe}, Eiji and {Kusakabe}, Nobuhiko and {Kwon}, Jungmi and {Ikoma}, Masahiro and {Hori}, Yasunori and {Genda}, Hidenori and {Fukui}, Akihiko and {Fujii}, Yuka and {Kawahara}, Hajime and {Olivier}, Guyon and {Jovanovic}, Nemanja and {Harakawa}, Hiroki and {Hayashi}, Masahiko and {Hidai}, Masahide and {Machida}, Masahiro and {Matsuo}, Taro and {Nagata}, Tetsuya and {Ogihara}, Masahiro and {Takami}, Hideki and {Takato}, Naruhisa and {Terada}, Hiroshi and {Oh}, Daehyeon},
        title = "{The infrared Doppler (IRD) instrument for the Subaru telescope: instrument description and commissioning results}",
    booktitle = {Ground-based and Airborne Instrumentation for Astronomy VII},
         year = 2018,
       editor = {{Evans}, Christopher J. and {Simard}, Luc and {Takami}, Hideki},
       series = {Society of Photo-Optical Instrumentation Engineers (SPIE) Conference Series},
       volume = {10702},
        month = jul,
          eid = {1070211},
        pages = {1070211},
          doi = {10.1117/12.2311836},
       adsurl = {https://ui.adsabs.harvard.edu/abs/2018SPIE10702E..11K}
}

@INPROCEEDINGS{Tamura2012,
       author = {{Tamura}, M. and {Suto}, H. and {Nishikawa}, J. and {Kotani}, T. and {Sato}, B. and {Aoki}, W. and {Usuda}, T. and {Kurokawa}, T. and {Kashiwagi}, K. and {Nishiyama}, S. and {Ikeda}, Y. and {Hall}, D. and {Hodapp}, K. and {Hashimoto}, J. and {Morino}, J. and {Inoue}, S. and {Mizuno}, Y. and {Washizaki}, Y. and {Tanaka}, Y. and {Suzuki}, S. and {Kwon}, J. and {Suenaga}, T. and {Oh}, D. and {Narita}, N. and {Kokubo}, E. and {Hayano}, Y. and {Izumiura}, H. and {Kambe}, E. and {Kudo}, T. and {Kusakabe}, N. and {Ikoma}, M. and {Hori}, Ya. and {Omiya}, M. and {Genda}, H. and {Fukui}, A. and {Fujii}, Y. and {Guyon}, O. and {Harakawa}, H. and {Hayashi}, M. and {Hidai}, M. and {Hirano}, T. and {Kuzuhara}, M. and {Machida}, M. and {Matsuo}, T. and {Nagata}, T. and {Ohnuki}, H. and {Ogihara}, M. and {Oshino}, S. and {Suzuki}, R. and {Takami}, H. and {Takato}, N. and {Takahashi}, Y. and {Tachinami}, C. and {Terada}, H.},
        title = "{Infrared Doppler instrument for the Subaru Telescope (IRD)}",
    booktitle = {Ground-based and Airborne Instrumentation for Astronomy IV},
         year = 2012,
       editor = {{McLean}, Ian S. and {Ramsay}, Suzanne K. and {Takami}, Hideki},
       series = {Society of Photo-Optical Instrumentation Engineers (SPIE) Conference Series},
       volume = {8446},
        month = sep,
          eid = {84461T},
        pages = {84461T},
          doi = {10.1117/12.925885},
       adsurl = {https://ui.adsabs.harvard.edu/abs/2012SPIE.8446E..1TT}
}

@ARTICLE{Garcia2022,
       author = {{Garcia}, Lionel J. and {Timmermans}, Mathilde and {Pozuelos}, Francisco J. and {Ducrot}, Elsa and {Gillon}, Micha{\"e}l and {Delrez}, Laetitia and {Wells}, Robert D. and {Jehin}, Emmanu{\"e}l},
        title = "{PROSE: a PYTHON framework for modular astronomical images processing}",
      journal = {\mnras},
         year = 2022,
        month = feb,
       volume = {509},
       number = {4},
        pages = {4817-4828},
          doi = {10.1093/mnras/stab3113},
archivePrefix = {arXiv},
       eprint = {2111.02814},
 primaryClass = {astro-ph.IM},
       adsurl = {https://ui.adsabs.harvard.edu/abs/2022MNRAS.509.4817G}
}

@ARTICLE{Barkaoui2019,
       author = {{Barkaoui}, K. and {Burdanov}, A. and {Hellier}, C. and {Gillon}, M. and {Smalley}, B. and {Maxted}, P.~F.~L. and {Lendl}, M. and {Triaud}, A.~H.~M.~J. and {Anderson}, D.~R. and {McCormac}, J. and {Jehin}, E. and {Almleaky}, Y. and {Armstrong}, D.~J. and {Benkhaldoun}, Z. and {Bouchy}, F. and {Brown}, D.~J.~A. and {Cameron}, A.~C. and {Daassou}, A. and {Delrez}, L. and {Ducrot}, E. and {Foxell}, E. and {Murray}, C. and {Nielsen}, L.~D. and {Pepe}, F. and {Pollacco}, D. and {Pozuelos}, F.~J. and {Queloz}, D. and {Segransan}, D. and {Udry}, S. and {Thompson}, S. and {West}, R.~G.},
        title = "{Discovery of Three New Transiting Hot Jupiters: WASP-161 b, WASP-163 b, and WASP-170 b}",
      journal = {\aj},
         year = 2019,
        month = feb,
       volume = {157},
       number = {2},
          eid = {43},
        pages = {43},
          doi = {10.3847/1538-3881/aaf422},
archivePrefix = {arXiv},
       eprint = {1807.06548},
 primaryClass = {astro-ph.EP},
       adsurl = {https://ui.adsabs.harvard.edu/abs/2019AJ....157...43B}
}

@ARTICLE{Gillon2013,
       author = {{Gillon}, M. and {Anderson}, D.~R. and {Collier-Cameron}, A. and {Doyle}, A.~P. and {Fumel}, A. and {Hellier}, C. and {Jehin}, E. and {Lendl}, M. and {Maxted}, P.~F.~L. and {Montalb{\'a}n}, J. and {Pepe}, F. and {Pollacco}, D. and {Queloz}, D. and {S{\'e}gransan}, D. and {Smith}, A.~M.~S. and {Smalley}, B. and {Southworth}, J. and {Triaud}, A.~H.~M.~J. and {Udry}, S. and {West}, R.~G.},
        title = "{WASP-64 b and WASP-72 b: two new transiting highly irradiated giant planets}",
      journal = {\aap},
         year = 2013,
        month = apr,
       volume = {552},
          eid = {A82},
        pages = {A82},
          doi = {10.1051/0004-6361/201220561},
archivePrefix = {arXiv},
       eprint = {1210.4257},
 primaryClass = {astro-ph.EP},
       adsurl = {https://ui.adsabs.harvard.edu/abs/2013A&A...552A..82G}
}

@ARTICLE{Jehin2011,
       author = {{Jehin}, E. and {Gillon}, M. and {Queloz}, D. and {Magain}, P. and {Manfroid}, J. and {Chantry}, V. and {Lendl}, M. and {Hutsem{\'e}kers}, D. and {Udry}, S.},
        title = "{TRAPPIST: TRAnsiting Planets and PlanetesImals Small Telescope}",
      journal = {The Messenger},
         year = 2011,
        month = sep,
       volume = {145},
        pages = {2-6},
       adsurl = {https://ui.adsabs.harvard.edu/abs/2011Msngr.145....2J}
}

@ARTICLE{Fukui2011,
       author = {{Fukui}, Akihiko and {Narita}, Norio and {Tristram}, Paul J. and {Sumi}, Takahiro and {Abe}, Fumio and {Itow}, Yoshitaka and {Sullivan}, Denis J. and {Bond}, Ian A. and {Hirano}, Teruyuki and {Tamura}, Motohide and {Bennett}, David P. and {Furusawa}, Kei and {Hayashi}, Fumiya and {Hearnshaw}, John B. and {Hosaka}, Shun and {Kamiya}, Koki and {Kobara}, Shuhei and {Korpela}, Aarno and {Kilmartin}, Pam M. and {Lin}, Wei and {Ling}, Cho Hong and {Makita}, Shota and {Masuda}, Kimiaki and {Matsubara}, Yutaka and {Miyake}, Noriyuki and {Muraki}, Yasushi and {Nagaya}, Maiko and {Nishimoto}, Kenta and {Ohnishi}, Kouji and {Omori}, Kengo and {Perrott}, Yvette and {Rattenbury}, Nicholas and {Saito}, Toshiharu and {Skuljan}, Ljiljana and {Suzuki}, Daisuke and {Sweatman}, Winston L. and {Wada}, Kohei},
        title = "{Measurements of Transit Timing Variations for WASP-5b}",
      journal = {\pasj},
         year = 2011,
        month = feb,
       volume = {63},
        pages = {287},
          doi = {10.1093/pasj/63.1.287},
archivePrefix = {arXiv},
       eprint = {1009.5769},
 primaryClass = {astro-ph.EP},
       adsurl = {https://ui.adsabs.harvard.edu/abs/2011PASJ...63..287F}
}

@INPROCEEDINGS{Narita2020,
       author = {{Narita}, Norio and {Fukui}, Akihiko and {Yamamuro}, Tomoyasu and {Harbeck}, Daniel and {Bowman}, Mark and {Elphick}, Mark and {Nation}, Jon and {Armstrong}, J.~D. and {Han}, Jacqueline and {Abe}, Shunichi and {Ikoma}, Masahiro and {Isogai}, Keisuke and {Kawauchi}, Kiyoe and {Kurita}, Seiya and {Kusakabe}, Nobuhiko and {de Leon}, Jerome and {Livingston}, John and {Mori}, Mayuko and {Nishiumi}, Taku and {Tamura}, Motohide and {Watanabe}, Noriharu and {Volgenau}, Nikolaus and {Heinrich-Josties}, Elisabeth and {Foale}, Steve and {Daily}, Matt and {McCully}, Curtis and {Kirby}, Annie and {Smith}, Cary and {Haworth}, Brian and {Conway}, Patrick and {Storrie-Lombardi}, Lisa and {Rosing}, Wayne and {Chatelain}, Joey and {Bachelet}, Etienne and {Johnson}, Marshall and {Rabus}, Markus},
        title = "{MuSCAT3: a 4-color simultaneous camera for the 2m Faulkes Telescope North}",
    booktitle = {Ground-based and Airborne Instrumentation for Astronomy VIII},
         year = 2020,
       editor = {{Evans}, Christopher J. and {Bryant}, Julia J. and {Motohara}, Kentaro},
       series = {Society of Photo-Optical Instrumentation Engineers (SPIE) Conference Series},
       volume = {11447},
        month = dec,
          eid = {114475K},
        pages = {114475K},
          doi = {10.1117/12.2559947},
       adsurl = {https://ui.adsabs.harvard.edu/abs/2020SPIE11447E..5KN}
}

@INPROCEEDINGS{McCully2018,
       author = {{McCully}, Curtis and {Volgenau}, Nikolaus H. and {Harbeck}, Daniel-Rolf and {Lister}, Tim A. and {Saunders}, Eric S. and {Turner}, Monica L. and {Siiverd}, Robert J. and {Bowman}, Mark},
        title = "{Real-time processing of the imaging data from the network of Las Cumbres Observatory Telescopes using BANZAI}",
    booktitle = {Software and Cyberinfrastructure for Astronomy V},
         year = 2018,
       editor = {{Guzman}, Juan C. and {Ibsen}, Jorge},
       series = {Society of Photo-Optical Instrumentation Engineers (SPIE) Conference Series},
       volume = {10707},
        month = jul,
          eid = {107070K},
        pages = {107070K},
          doi = {10.1117/12.2314340},
archivePrefix = {arXiv},
       eprint = {1811.04163},
 primaryClass = {astro-ph.IM},
       adsurl = {https://ui.adsabs.harvard.edu/abs/2018SPIE10707E..0KM}
}

@INPROCEEDINGS{Collins2019,
       author = {{Collins}, Karen},
        title = "{TESS Follow-up Observing Program Working Group (TFOP WG) Sub Group 1 (SG1): Ground-based Time-series Photometry}",
    booktitle = {American Astronomical Society Meeting Abstracts \#233},
         year = 2019,
       series = {American Astronomical Society Meeting Abstracts},
       volume = {233},
        month = jan,
          eid = {140.05},
        pages = {140.05},
       adsurl = {https://ui.adsabs.harvard.edu/abs/2019AAS...23314005C}
}

@ARTICLE{espinoza2016,
       author = {{Espinoza}, N. and {Bayliss}, D. and {Hartman}, J.~D. and {Bakos}, G. {\'A}. and {Jord{\'a}n}, A. and {Zhou}, G. and {Mancini}, L. and {Brahm}, R. and {Ciceri}, S. and {Bhatti}, W. and {Csubry}, Z. and {Rabus}, M. and {Penev}, K. and {Bento}, J. and {de Val-Borro}, M. and {Henning}, T. and {Schmidt}, B. and {Suc}, V. and {Wright}, D.~J. and {Tinney}, C.~G. and {Tan}, T.~G. and {Noyes}, R.},
        title = "{HATS-25b through HATS-30b: A Half-dozen New Inflated Transiting Hot Jupiters from the HATSouth Survey}",
      journal = {\aj},
         year = 2016,
        month = oct,
       volume = {152},
       number = {4},
          eid = {108},
        pages = {108},
          doi = {10.3847/0004-6256/152/4/108},
archivePrefix = {arXiv},
       eprint = {1606.00023},
 primaryClass = {astro-ph.EP},
       adsurl = {https://ui.adsabs.harvard.edu/abs/2016AJ....152..108E}
}

@ARTICLE{furlan2017,
       author = {{Furlan}, E. and {Ciardi}, D.~R. and {Everett}, M.~E. and {Saylors}, M. and {Teske}, J.~K. and {Horch}, E.~P. and {Howell}, S.~B. and {van Belle}, G.~T. and {Hirsch}, L.~A. and {Gautier}, III, T.~N. and {Adams}, E.~R. and {Barrado}, D. and {Cartier}, K.~M.~S. and {Dressing}, C.~D. and {Dupree}, A.~K. and {Gilliland}, R.~L. and {Lillo-Box}, J. and {Lucas}, P.~W. and {Wang}, J.},
        title = "{The Kepler Follow-up Observation Program. I. A Catalog of Companions to Kepler Stars from High-Resolution Imaging}",
      journal = {\aj},
         year = 2017,
        month = feb,
       volume = {153},
       number = {2},
          eid = {71},
        pages = {71},
          doi = {10.3847/1538-3881/153/2/71},
archivePrefix = {arXiv},
       eprint = {1612.02392},
 primaryClass = {astro-ph.SR},
       adsurl = {https://ui.adsabs.harvard.edu/abs/2017AJ....153...71F}
}

@INPROCEEDINGS{srinath2014,
       author = {{Srinath}, Srikar and {McGurk}, Rosalie and {Rockosi}, Constance and {Kupke}, Renate and {Gavel}, Donald and {Cabak}, Gerald and {Cowley}, David and {Peck}, Michael and {Ratliff}, Christopher and {Gates}, Elinor and {Dillon}, Daren and {Norton}, Andrew and {Reining}, Marc},
        title = "{Swimming with ShARCS: comparison of on-sky sensitivity with model predictions for ShaneAO on the Lick Observatory 3-meter telescope}",
    booktitle = {Adaptive Optics Systems IV},
         year = 2014,
       editor = {{Marchetti}, Enrico and {Close}, Laird M. and {Vran}, Jean-Pierre},
       series = {Society of Photo-Optical Instrumentation Engineers (SPIE) Conference Series},
       volume = {9148},
        month = aug,
          eid = {91482Z},
        pages = {91482Z},
          doi = {10.1117/12.2055672},
archivePrefix = {arXiv},
       eprint = {1407.8206},
 primaryClass = {astro-ph.IM},
       adsurl = {https://ui.adsabs.harvard.edu/abs/2014SPIE.9148E..2ZS}
}

@article{clough2005atmospheric,
  title={Atmospheric radiative transfer modeling: A summary of the AER codes},
  author={Clough, Shepard A and Shephard, Mark W and Mlawer, Eli J and Delamere, Jennifer S and Iacono, Michael J and Cady-Pereira, Karen and Boukabara, Sid and Brown, Patrick D},
  journal={Journal of Quantitative Spectroscopy and Radiative Transfer},
  volume={91},
  number={2},
  pages={233--244},
  year={2005},
  publisher={Elsevier}
}

@ARTICLE{gullikson2014,
       author = {{Gullikson}, Kevin and {Dodson-Robinson}, Sarah and {Kraus}, Adam},
        title = "{Correcting for Telluric Absorption: Methods, Case Studies, and Release of the TelFit Code}",
      journal = {\aj},
         year = 2014,
        month = sep,
       volume = {148},
       number = {3},
          eid = {53},
        pages = {53},
          doi = {10.1088/0004-6256/148/3/53},
archivePrefix = {arXiv},
       eprint = {1406.6059},
 primaryClass = {astro-ph.IM},
       adsurl = {https://ui.adsabs.harvard.edu/abs/2014AJ....148...53G}
}

@ARTICLE{anglada_escude2012,
       author = {{Anglada-Escud{\'e}}, Guillem and {Butler}, R. Paul},
        title = "{The HARPS-TERRA Project. I. Description of the Algorithms, Performance, and New Measurements on a Few Remarkable Stars Observed by HARPS}",
      journal = {\apjs},
         year = 2012,
        month = jun,
       volume = {200},
       number = {2},
          eid = {15},
        pages = {15},
          doi = {10.1088/0067-0049/200/2/15},
archivePrefix = {arXiv},
       eprint = {1202.2570},
 primaryClass = {astro-ph.EP},
       adsurl = {https://ui.adsabs.harvard.edu/abs/2012ApJS..200...15A}
}

@ARTICLE{shetrone2007,
       author = {{Shetrone}, Matthew and {Cornell}, Mark E. and {Fowler}, James R. and {Gaffney}, Niall and {Laws}, Benjamin and {Mader}, Jeff and {Mason}, Cloud and {Odewahn}, Stephen and {Roman}, Brian and {Rostopchin}, Sergey and {Schneider}, Donald P. and {Umbarger}, James and {Westfall}, Amy},
        title = "{Ten Year Review of Queue Scheduling of the Hobby-Eberly Telescope}",
      journal = {\pasp},
         year = 2007,
        month = may,
       volume = {119},
       number = {855},
        pages = {556-566},
          doi = {10.1086/519291},
archivePrefix = {arXiv},
       eprint = {0705.3889},
 primaryClass = {astro-ph},
       adsurl = {https://ui.adsabs.harvard.edu/abs/2007PASP..119..556S}
}

@ARTICLE{metcalf2019,
       author = {{Metcalf}, Andrew J. and {Anderson}, Tyler and {Bender}, Chad F. and {Blakeslee}, Scott and {Brand}, Wesley and {Carlson}, David R. and {Cochran}, William D. and {Diddams}, Scott A. and {Endl}, Michael and {Fredrick}, Connor and {Halverson}, Sam and {Hickstein}, Daniel D. and {Hearty}, Fred and {Jennings}, Jeff and {Kanodia}, Shubham and {Kaplan}, Kyle F. and {Levi}, Eric and {Lubar}, Emily and {Mahadevan}, Suvrath and {Monson}, Andrew and {Ninan}, Joe P. and {Nitroy}, Colin and {Osterman}, Steve and {Papp}, Scott B. and {Quinlan}, Franklyn and {Ramsey}, Larry and {Robertson}, Paul and {Roy}, Arpita and {Schwab}, Christian and {Sigurdsson}, Steinn and {Srinivasan}, Kartik and {Stefansson}, Gudmundur and {Sterner}, David A. and {Terrien}, Ryan and {Wolszczan}, Alex and {Wright}, Jason T. and {Ycas}, Gabriel},
        title = "{Stellar spectroscopy in the near-infrared with a laser frequency comb}",
      journal = {Optica},
         year = 2019,
        month = feb,
       volume = {6},
       number = {2},
        pages = {233},
          doi = {10.1364/OPTICA.6.000233},
archivePrefix = {arXiv},
       eprint = {1902.00500},
 primaryClass = {astro-ph.EP},
       adsurl = {https://ui.adsabs.harvard.edu/abs/2019Optic...6..233M}
}

@ARTICLE{stefansson2016,
       author = {{Stefansson}, Gudmundur and {Hearty}, Frederick and {Robertson}, Paul and {Mahadevan}, Suvrath and {Anderson}, Tyler and {Levi}, Eric and {Bender}, Chad and {Nelson}, Matthew and {Monson}, Andrew and {Blank}, Basil and {Halverson}, Samuel and {Henderson}, Chuck and {Ramsey}, Lawrence and {Roy}, Arpita and {Schwab}, Christian and {Terrien}, Ryan},
        title = "{A Versatile Technique to Enable Sub-milli-Kelvin Instrument Stability for Precise Radial Velocity Measurements: Tests with the Habitable-zone Planet Finder}",
      journal = {\apj},
         year = 2016,
        month = dec,
       volume = {833},
       number = {2},
          eid = {175},
        pages = {175},
          doi = {10.3847/1538-4357/833/2/175},
archivePrefix = {arXiv},
       eprint = {1610.06216},
 primaryClass = {astro-ph.IM},
       adsurl = {https://ui.adsabs.harvard.edu/abs/2016ApJ...833..175S}
}

@ARTICLE{Zhao2014,
       author = {{Zhao}, Ming and {O'Rourke}, Joseph G. and {Wright}, Jason T. and {Knutson}, Heather A. and {Burrows}, Adam and {Fortney}, Johnathan and {Ngo}, Henry and {Fulton}, Benjamin J. and {Baranec}, Christoph and {Riddle}, Reed and {Law}, Nicholas M. and {Muirhead}, Philip S. and {Hinkley}, Sasha and {Showman}, Adam P. and {Curtis}, Jason and {Burruss}, Rick},
        title = "{Characterization of the Atmosphere of the Hot Jupiter HAT-P-32Ab and the M-dwarf Companion HAT-P-32B}",
      journal = {\apj},
         year = 2014,
        month = dec,
       volume = {796},
       number = {2},
          eid = {115},
        pages = {115},
          doi = {10.1088/0004-637X/796/2/115},
archivePrefix = {arXiv},
       eprint = {1410.0968},
 primaryClass = {astro-ph.EP},
       adsurl = {https://ui.adsabs.harvard.edu/abs/2014ApJ...796..115Z}
}

@ARTICLE{Stefansson2018b,
       author = {{Stefansson}, Gudmundur and {Li}, Yiting and {Mahadevan}, Suvrath and {Wisniewski}, John and {Hebb}, Leslie and {Morris}, Brett and {Huehnerhoff}, Joseph and {Hawley}, Suzanne},
        title = "{Diffuser-assisted Photometric Follow-up Observations of the Neptune-sized Planets K2-28b and K2-100b}",
      journal = {\aj},
         year = 2018,
        month = dec,
       volume = {156},
       number = {6},
          eid = {266},
        pages = {266},
          doi = {10.3847/1538-3881/aae6ca},
archivePrefix = {arXiv},
       eprint = {1807.04420},
 primaryClass = {astro-ph.EP},
       adsurl = {https://ui.adsabs.harvard.edu/abs/2018AJ....156..266S}
}

@INPROCEEDINGS{Huehnerhoff2016,
       author = {{Huehnerhoff}, Joseph and {Ketzeback}, William and {Bradley}, Alaina and {Dembicky}, Jack and {Doughty}, Caitlin and {Hawley}, Suzanne and {Johnson}, Courtney and {Klaene}, Mark and {Leon}, Ed and {McMillan}, Russet and {Owen}, Russell and {Sayres}, Conor and {Sheen}, Tyler and {Shugart}, Alysha},
        title = "{Astrophysical Research Consortium Telescope Imaging Camera (ARCTIC) facility optical imager for the Apache Point Observatory 3.5m telescope}",
    booktitle = {Ground-based and Airborne Instrumentation for Astronomy VI},
         year = 2016,
       editor = {{Evans}, Christopher J. and {Simard}, Luc and {Takami}, Hideki},
       series = {Society of Photo-Optical Instrumentation Engineers (SPIE) Conference Series},
       volume = {9908},
        month = aug,
          eid = {99085H},
        pages = {99085H},
          doi = {10.1117/12.2234214},
       adsurl = {https://ui.adsabs.harvard.edu/abs/2016SPIE.9908E..5HH}
}

@INPROCEEDINGS{stefansson2018a,
       author = {{Stefansson}, Gudmundur and {Mahadevan}, Suvrath and {Wisniewski}, John and {Li}, Yiting and {Hebb}, Leslie and {Morris}, Brett and {Halverson}, Samuel and {Monson}, Andrew and {Robertson}, Paul},
        title = "{Extreme precision photometry from the ground with beam-shaping diffusers for K2, TESS, and beyond}",
    booktitle = {Ground-based and Airborne Instrumentation for Astronomy VII},
         year = 2018,
       editor = {{Evans}, Christopher J. and {Simard}, Luc and {Takami}, Hideki},
       series = {Society of Photo-Optical Instrumentation Engineers (SPIE) Conference Series},
       volume = {10702},
        month = aug,
          eid = {1070250},
        pages = {1070250},
          doi = {10.1117/12.2312833},
archivePrefix = {arXiv},
       eprint = {1808.02187},
 primaryClass = {astro-ph.IM},
       adsurl = {https://ui.adsabs.harvard.edu/abs/2018SPIE10702E..50S}
}

@ARTICLE{stefansson2017,
       author = {{Stefansson}, Gudmundur and {Mahadevan}, Suvrath and {Hebb}, Leslie and {Wisniewski}, John and {Huehnerhoff}, Joseph and {Morris}, Brett and {Halverson}, Sam and {Zhao}, Ming and {Wright}, Jason and {O'rourke}, Joseph and {Knutson}, Heather and {Hawley}, Suzanne and {Kanodia}, Shubham and {Li}, Yiting and {Hagen}, Lea M.~Z. and {Liu}, Leo J. and {Beatty}, Thomas and {Bender}, Chad and {Robertson}, Paul and {Dembicky}, Jack and {Gray}, Candace and {Ketzeback}, William and {McMillan}, Russet and {Rudyk}, Theodore},
        title = "{Toward Space-like Photometric Precision from the Ground with Beam-shaping Diffusers}",
      journal = {\apj},
         year = 2017,
        month = oct,
       volume = {848},
       number = {1},
          eid = {9},
        pages = {9},
          doi = {10.3847/1538-4357/aa88aa},
archivePrefix = {arXiv},
       eprint = {1710.01790},
 primaryClass = {astro-ph.IM},
       adsurl = {https://ui.adsabs.harvard.edu/abs/2017ApJ...848....9S}
}

@ARTICLE{Collins2017,
       author = {{Collins}, Karen A. and {Kielkopf}, John F. and {Stassun}, Keivan G. and {Hessman}, Frederic V.},
        title = "{AstroImageJ: Image Processing and Photometric Extraction for Ultra-precise Astronomical Light Curves}",
      journal = {\aj},
         year = 2017,
        month = feb,
       volume = {153},
       number = {2},
          eid = {77},
        pages = {77},
          doi = {10.3847/1538-3881/153/2/77},
archivePrefix = {arXiv},
       eprint = {1701.04817},
 primaryClass = {astro-ph.IM},
       adsurl = {https://ui.adsabs.harvard.edu/abs/2017AJ....153...77C}
}

@ARTICLE{Hori2024,
       author = {{Hori}, Yasunori and {Fukui}, Akihiko and {Hirano}, Teruyuki and {Narita}, Norio and {de Leon}, Jerome P. and {Ishikawa}, Hiroyuki Tako and {Hartman}, Joel D. and {Morello}, Giuseppe and {Garc{\'\i}a}, Nestor Abreu and {{\'A}lvarez Hern{\'a}ndez}, Leticia and {B{\'e}jar}, V{\'\i}ctor J.~S. and {Calatayud-Borras}, Y{\'e}ssica and {Carleo}, Ilaria and {Enoc}, Gareb and {Esparza-Borges}, Emma and {Fukuda}, Izuru and {Gal{\'a}n}, Daniel and {Gerald{\'\i}a-Gonz{\'a}lez}, Samuel and {Hayashi}, Yuya and {Ikoma}, Masahiro and {Ikuta}, Kai and {Isogai}, Keisuke and {Kagetani}, Taiki and {Kawai}, Yugo and {Kawauchi}, Kiyoe and {Kimura}, Tadahiro and {Kodama}, Takanori and {Korth}, Judith and {Kusakabe}, Nobuhiko and {Laza-Ramos}, Andr{\'e}s and {Livingston}, John H. and {Luque}, Rafael and {Miyakawa}, Kohei and {Mori}, Mayuko and {Murgas}, Felipe and {Orell-Miquel}, Jaume and {Palle}, Enric and {Parviainen}, Hannu and {Pel{\'a}ez-Torres}, Alberto and {Puig-Subir{\`a}}, Marta and {S{\'a}nchez-Benavente}, Manuel and {Sosa-Guill{\'e}n}, Paula and {Stangret}, Monika and {Terada}, Yuka and {Mu{\~n}oz Torres}, Sara and {Watanabe}, Noriharu and {Bakos}, Gaspar {\'A}. and {Barkaoui}, Khalid and {Beichman}, Charles and {Benkhaldoun}, Zouhair and {Boyle}, Andrew W. and {Ciardi}, David R. and {Clark}, Catherine A. and {Collins}, Karen A. and {Collins}, Kevin I. and {Conti}, Dennis M. and {Crossfield}, Ian J.~M. and {Everett}, Mark E. and {Furlan}, Elise and {Ghachoui}, Mourad and {Gillon}, Micha{\"e}l and {Gonzales}, Erica J. and {Higuera}, Jesus and {Horne}, Keith and {Howell}, Steve B. and {Jehin}, Emmanu{\"e}l and {Lester}, Kathryn V. and {Lund}, Michael B. and {Matson}, Rachel and {Matthews}, Elisabeth C. and {Pozuelos}, Francisco J. and {Safonov}, Boris S. and {Schlieder}, Joshua E. and {Schwarz}, Richard P. and {Sefako}, Ramotholo and {Srdoc}, Gregor and {Strakhov}, Ivan A. and {Timmermans}, Mathilde and {Waalkes}, William C. and {Ziegler}, Carl and {Charbonneau}, David and {Essack}, Zahra and {Guerrero}, Natalia M. and {Harakawa}, Hiroki and {Hedges}, Christina and {Ishizuka}, Masato and {Jenkins}, Jon M. and {Konishi}, Mihoko and {Kotani}, Takayuki and {Kudo}, Tomoyuki and {Kurokawa}, Takashi and {Kuzuhara}, Masayuki and {Nishikawa}, Jun and {Omiya}, Masashi and {Ricker}, George R. and {Seager}, Sara and {Serizawa}, Takuma and {Striegel}, Stephanie and {Tamura}, Motohide and {Ueda}, Akitoshi and {Vanderspek}, Roland and {Vievard}, S{\'e}bastien and {Winn}, Joshua N.},
        title = "{The Discovery and Follow-up of Four Transiting Short-period Sub-Neptunes Orbiting M Dwarfs}",
      journal = {\aj},
         year = 2024,
        month = jun,
       volume = {167},
       number = {6},
          eid = {289},
        pages = {289},
          doi = {10.3847/1538-3881/ad4115},
archivePrefix = {arXiv},
       eprint = {2405.12637},
 primaryClass = {astro-ph.EP},
       adsurl = {https://ui.adsabs.harvard.edu/abs/2024AJ....167..289H}
}

@INPROCEEDINGS{Jenkins2016,
       author = {{Jenkins}, Jon M. and {Twicken}, Joseph D. and {McCauliff}, Sean and {Campbell}, Jennifer and {Sanderfer}, Dwight and {Lung}, David and {Mansouri-Samani}, Masoud and {Girouard}, Forrest and {Tenenbaum}, Peter and {Klaus}, Todd and {Smith}, Jeffrey C. and {Caldwell}, Douglas A. and {Chacon}, A.~D. and {Henze}, Christopher and {Heiges}, Cory and {Latham}, David W. and {Morgan}, Edward and {Swade}, Daryl and {Rinehart}, Stephen and {Vanderspek}, Roland},
        title = "{The TESS science processing operations center}",
    booktitle = {Software and Cyberinfrastructure for Astronomy IV},
         year = 2016,
       editor = {{Chiozzi}, Gianluca and {Guzman}, Juan C.},
       series = {Society of Photo-Optical Instrumentation Engineers (SPIE) Conference Series},
       volume = {9913},
        month = aug,
          eid = {99133E},
        pages = {99133E},
          doi = {10.1117/12.2233418},
       adsurl = {https://ui.adsabs.harvard.edu/abs/2016SPIE.9913E..3EJ}
}

@ARTICLE{Ford2006,
       author = {{Ford}, Eric B.},
        title = "{Improving the Efficiency of Markov Chain Monte Carlo for Analyzing the Orbits of Extrasolar Planets}",
      journal = {\apj},
         year = 2006,
        month = may,
       volume = {642},
       number = {1},
        pages = {505-522},
          doi = {10.1086/500802},
archivePrefix = {arXiv},
       eprint = {astro-ph/0512634},
 primaryClass = {astro-ph},
       adsurl = {https://ui.adsabs.harvard.edu/abs/2006ApJ...642..505F}
}

@ARTICLE{Stefansson2023,
       author = {{Stef{\'a}nsson}, Gu{\dh}mundur and {Mahadevan}, Suvrath and {Miguel}, Yamila and {Robertson}, Paul and {Delamer}, Megan and {Kanodia}, Shubham and {Ca{\~n}as}, Caleb I. and {Winn}, Joshua N. and {Ninan}, Joe P. and {Terrien}, Ryan C. and {Holcomb}, Rae and {Ford}, Eric B. and {Zawadzki}, Brianna and {Bowler}, Brendan P. and {Bender}, Chad F. and {Cochran}, William D. and {Diddams}, Scott and {Endl}, Michael and {Fredrick}, Connor and {Halverson}, Samuel and {Hearty}, Fred and {Hill}, Gary J. and {Lin}, Andrea S.~J. and {Metcalf}, Andrew J. and {Monson}, Andrew and {Ramsey}, Lawrence and {Roy}, Arpita and {Schwab}, Christian and {Wright}, Jason T. and {Zeimann}, Gregory},
        title = "{A Neptune-mass exoplanet in close orbit around a very low-mass star challenges formation models}",
      journal = {Science},
         year = 2023,
        month = dec,
       volume = {382},
       number = {6674},
        pages = {1031-1035},
          doi = {10.1126/science.abo0233},
archivePrefix = {arXiv},
       eprint = {2303.13321},
 primaryClass = {astro-ph.EP},
       adsurl = {https://ui.adsabs.harvard.edu/abs/2023Sci...382.1031S}
}

@ARTICLE{Hill2021,
       author = {{Hill}, Gary J. and {Lee}, Hanshin and {MacQueen}, Phillip J. and {Kelz}, Andreas and {Drory}, Niv and {Vattiat}, Brian L. and {Good}, John M. and {Ramsey}, Jason and {Kriel}, Herman and {Peterson}, Trent and {DePoy}, D.~L. and {Gebhardt}, Karl and {Marshall}, J.~L. and {Tuttle}, Sarah E. and {Bauer}, Svend M. and {Chonis}, Taylor S. and {Fabricius}, Maximilian H. and {Froning}, Cynthia and {H{\"a}user}, Marco and {Indahl}, Briana L. and {Jahn}, Thomas and {Landriau}, Martin and {Leck}, Ron and {Montesano}, Francesco and {Prochaska}, Travis and {Snigula}, Jan M. and {Zeimann}, Greg and {Bryant}, Randy and {Damm}, George and {Fowler}, J.~R. and {Janowiecki}, Steven and {Martin}, Jerry and {Mrozinski}, Emily and {Odewahn}, Stephen and {Rostopchin}, Sergey and {Shetrone}, Matthew and {Spencer}, Renny and {Mentuch Cooper}, Erin and {Armandroff}, Taft and {Bender}, Ralf and {Dalton}, Gavin and {Hopp}, Ulrich and {Komatsu}, Eiichiro and {Nicklas}, Harald and {Ramsey}, Lawrence W. and {Roth}, Martin M. and {Schneider}, Donald P. and {Sneden}, Chris and {Steinmetz}, Matthias},
        title = "{The HETDEX Instrumentation: Hobby-Eberly Telescope Wide-field Upgrade and VIRUS}",
      journal = {\aj},
         year = 2021,
        month = dec,
       volume = {162},
       number = {6},
          eid = {298},
        pages = {298},
          doi = {10.3847/1538-3881/ac2c02},
archivePrefix = {arXiv},
       eprint = {2110.03843},
 primaryClass = {astro-ph.IM},
       adsurl = {https://ui.adsabs.harvard.edu/abs/2021AJ....162..298H}
}

@article{salvatier2016probabilistic,
  title={Probabilistic programming in Python using PyMC3},
  author={Salvatier, John and Wiecki, Thomas V and Fonnesbeck, Christopher},
  journal={PeerJ Computer Science},
  volume={2},
  pages={e55},
  year={2016},
  publisher={PeerJ Inc.}
}

@ARTICLE{hxrgproc,
       author = {{Rauscher}, Bernard J.},
        title = "{Teledyne H1RG, H2RG, and H4RG Noise Generator}",
      journal = {\pasp},
         year = 2015,
        month = nov,
       volume = {127},
       number = {957},
        pages = {1144},
          doi = {10.1086/684082},
archivePrefix = {arXiv},
       eprint = {1509.06264},
 primaryClass = {astro-ph.IM},
       adsurl = {https://ui.adsabs.harvard.edu/abs/2015PASP..127.1144R}
}

@misc{exoplanet_dfm,
       author = {{Foreman-Mackey}, Dan and {Barentsen}, Geert and {Barclay}, Tom},
        title = "{dfm/exoplanet: exoplanet v0.1.5}",
         year = 2019,
        month = mar,
          eid = {10.5281/zenodo.2587222},
          doi = {10.5281/zenodo.2587222},
      version = {v0.1.5},
    publisher = {Zenodo},
       adsurl = {https://ui.adsabs.harvard.edu/abs/2019zndo...2587222F}
}

@ARTICLE{astroimagej,
       author = {{Collins}, Karen A. and {Kielkopf}, John F. and {Stassun}, Keivan G. and {Hessman}, Frederic V.},
        title = "{AstroImageJ: Image Processing and Photometric Extraction for Ultra-precise Astronomical Light Curves}",
      journal = {\aj},
         year = 2017,
        month = feb,
       volume = {153},
       number = {2},
          eid = {77},
        pages = {77},
          doi = {10.3847/1538-3881/153/2/77},
archivePrefix = {arXiv},
       eprint = {1701.04817},
 primaryClass = {astro-ph.IM},
       adsurl = {https://ui.adsabs.harvard.edu/abs/2017AJ....153...77C}
}

@ARTICLE{ricker2015_tess,
       author = {{Ricker}, George R. and {Winn}, Joshua N. and {Vanderspek}, Roland and {Latham}, David W. and {Bakos}, G{\'a}sp{\'a}r {\'A}. and {Bean}, Jacob L. and {Berta-Thompson}, Zachory K. and {Brown}, Timothy M. and {Buchhave}, Lars and {Butler}, Nathaniel R. and {Butler}, R. Paul and {Chaplin}, William J. and {Charbonneau}, David and {Christensen-Dalsgaard}, J{\o}rgen and {Clampin}, Mark and {Deming}, Drake and {Doty}, John and {De Lee}, Nathan and {Dressing}, Courtney and {Dunham}, Edward W. and {Endl}, Michael and {Fressin}, Francois and {Ge}, Jian and {Henning}, Thomas and {Holman}, Matthew J. and {Howard}, Andrew W. and {Ida}, Shigeru and {Jenkins}, Jon M. and {Jernigan}, Garrett and {Johnson}, John Asher and {Kaltenegger}, Lisa and {Kawai}, Nobuyuki and {Kjeldsen}, Hans and {Laughlin}, Gregory and {Levine}, Alan M. and {Lin}, Douglas and {Lissauer}, Jack J. and {MacQueen}, Phillip and {Marcy}, Geoffrey and {McCullough}, Peter R. and {Morton}, Timothy D. and {Narita}, Norio and {Paegert}, Martin and {Palle}, Enric and {Pepe}, Francesco and {Pepper}, Joshua and {Quirrenbach}, Andreas and {Rinehart}, Stephen A. and {Sasselov}, Dimitar and {Sato}, Bun'ei and {Seager}, Sara and {Sozzetti}, Alessandro and {Stassun}, Keivan G. and {Sullivan}, Peter and {Szentgyorgyi}, Andrew and {Torres}, Guillermo and {Udry}, Stephane and {Villasenor}, Joel},
        title = "{Transiting Exoplanet Survey Satellite (TESS)}",
      journal = {Journal of Astronomical Telescopes, Instruments, and Systems},
         year = 2015,
        month = jan,
       volume = {1},
          eid = {014003},
        pages = {014003},
          doi = {10.1117/1.JATIS.1.1.014003},
       adsurl = {https://ui.adsabs.harvard.edu/abs/2015JATIS...1a4003R}
}

@ARTICLE{Howell2011,
       author = {{Howell}, Steve B. and {Everett}, Mark E. and {Sherry}, William and {Horch}, Elliott and {Ciardi}, David R.},
        title = "{Speckle Camera Observations for the NASA Kepler Mission Follow-up Program}",
      journal = {\aj},
         year = 2011,
        month = jul,
       volume = {142},
       number = {1},
          eid = {19},
        pages = {19},
          doi = {10.1088/0004-6256/142/1/19},
       adsurl = {https://ui.adsabs.harvard.edu/abs/2011AJ....142...19H}
}

@ARTICLE{stassun2018_tic,
       author = {{Stassun}, Keivan G. and {Oelkers}, Ryan J. and {Pepper}, Joshua and {Paegert}, Martin and {De Lee}, Nathan and {Torres}, Guillermo and {Latham}, David W. and {Charpinet}, St{\'e}phane and {Dressing}, Courtney D. and {Huber}, Daniel and {Kane}, Stephen R. and {L{\'e}pine}, S{\'e}bastien and {Mann}, Andrew and {Muirhead}, Philip S. and {Rojas-Ayala}, B{\'a}rbara and {Silvotti}, Roberto and {Fleming}, Scott W. and {Levine}, Al and {Plavchan}, Peter},
        title = "{The TESS Input Catalog and Candidate Target List}",
      journal = {\aj},
         year = 2018,
        month = sep,
       volume = {156},
       number = {3},
          eid = {102},
        pages = {102},
          doi = {10.3847/1538-3881/aad050},
archivePrefix = {arXiv},
       eprint = {1706.00495},
 primaryClass = {astro-ph.EP},
       adsurl = {https://ui.adsabs.harvard.edu/abs/2018AJ....156..102S}
}

@misc{cutri2003_2mass,
       author = {{Cutri}, R.~M. and {Skrutskie}, M.~F. and {van Dyk}, S. and {Beichman}, C.~A. and {Carpenter}, J.~M. and {Chester}, T. and {Cambresy}, L. and {Evans}, T. and {Fowler}, J. and {Gizis}, J. and {Howard}, E. and {Huchra}, J. and {Jarrett}, T. and {Kopan}, E.~L. and {Kirkpatrick}, J.~D. and {Light}, R.~M. and {Marsh}, K.~A. and {McCallon}, H. and {Schneider}, S. and {Stiening}, R. and {Sykes}, M. and {Weinberg}, M. and {Wheaton}, W.~A. and {Wheelock}, S. and {Zacarias}, N.},
        title = "{VizieR Online Data Catalog: 2MASS All-Sky Catalog of Point Sources (Cutri+ 2003)}",
 howpublished = {VizieR On-line Data Catalog: II/246.  Originally published in: 2003yCat.2246....0C},
         year = 2003,
        month = jun,
          eid = {II/246},
       adsurl = {https://ui.adsabs.harvard.edu/abs/2003yCat.2246....0C}
}

@INPROCEEDINGS{henden2018_apass,
       author = {{Henden}, Arne A. and {Levine}, Stephen and {Terrell}, Dirk and {Welch}, Douglas L. and {Munari}, Ulisse and {Kloppenborg}, Brian K.},
        title = "{APASS Data Release 10}",
    booktitle = {American Astronomical Society Meeting Abstracts \#232},
         year = 2018,
       series = {American Astronomical Society Meeting Abstracts},
       volume = {232},
        month = jun,
          eid = {223.06},
        pages = {223.06},
       adsurl = {https://ui.adsabs.harvard.edu/abs/2018AAS...23222306H}
}

@ARTICLE{kipping2023a,
       author = {{Kipping}, David M.},
        title = "{Efficient, uninformative sampling of limb darkening coefficients for two-parameter laws}",
      journal = {\mnras},
         year = 2013,
        month = nov,
       volume = {435},
       number = {3},
        pages = {2152-2160},
          doi = {10.1093/mnras/stt1435},
archivePrefix = {arXiv},
       eprint = {1308.0009},
 primaryClass = {astro-ph.SR},
       adsurl = {https://ui.adsabs.harvard.edu/abs/2013MNRAS.435.2152K}
}

@ARTICLE{mandel_agol2002,
       author = {{Mandel}, Kaisey and {Agol}, Eric},
        title = "{Analytic Light Curves for Planetary Transit Searches}",
      journal = {\apjl},
         year = 2002,
        month = dec,
       volume = {580},
       number = {2},
        pages = {L171-L175},
          doi = {10.1086/345520},
archivePrefix = {arXiv},
       eprint = {astro-ph/0210099},
 primaryClass = {astro-ph},
       adsurl = {https://ui.adsabs.harvard.edu/abs/2002ApJ...580L.171M}
}

@ARTICLE{agol2020,
       author = {{Agol}, Eric and {Luger}, Rodrigo and {Foreman-Mackey}, Daniel},
        title = "{Analytic Planetary Transit Light Curves and Derivatives for Stars with Polynomial Limb Darkening}",
      journal = {\aj},
         year = 2020,
        month = mar,
       volume = {159},
       number = {3},
          eid = {123},
        pages = {123},
          doi = {10.3847/1538-3881/ab4fee},
archivePrefix = {arXiv},
       eprint = {1908.03222},
 primaryClass = {astro-ph.EP},
       adsurl = {https://ui.adsabs.harvard.edu/abs/2020AJ....159..123A}
}

@ARTICLE{luger2019,
       author = {{Luger}, Rodrigo and {Agol}, Eric and {Foreman-Mackey}, Daniel and {Fleming}, David P. and {Lustig-Yaeger}, Jacob and {Deitrick}, Russell},
        title = "{starry: Analytic Occultation Light Curves}",
      journal = {\aj},
         year = 2019,
        month = feb,
       volume = {157},
       number = {2},
          eid = {64},
        pages = {64},
          doi = {10.3847/1538-3881/aae8e5},
archivePrefix = {arXiv},
       eprint = {1810.06559},
 primaryClass = {astro-ph.IM},
       adsurl = {https://ui.adsabs.harvard.edu/abs/2019AJ....157...64L}
}

@ARTICLE{bailerjones2021_gdr3,
       author = {{Bailer-Jones}, C.~A.~L. and {Rybizki}, J. and {Fouesneau}, M. and {Demleitner}, M. and {Andrae}, R.},
        title = "{Estimating Distances from Parallaxes. V. Geometric and Photogeometric Distances to 1.47 Billion Stars in Gaia Early Data Release 3}",
      journal = {\aj},
         year = 2021,
        month = mar,
       volume = {161},
       number = {3},
          eid = {147},
        pages = {147},
          doi = {10.3847/1538-3881/abd806},
archivePrefix = {arXiv},
       eprint = {2012.05220},
 primaryClass = {astro-ph.SR},
       adsurl = {https://ui.adsabs.harvard.edu/abs/2021AJ....161..147B}
}

@article{wright_eastman_2014,
	title = {Barycentric {Corrections} at 1 cm/s for precise {Doppler} velocities},
	volume = {126},
	issn = {00046280, 15383873},
	url = {http://arxiv.org/abs/1409.4774},
	doi = {10.1086/678541},
	number = {943},
	urldate = {2018-04-29},
	journal = {Publications of the Astronomical Society of the Pacific},
	author = {Wright, J. T. and Eastman, J. D.},
	month = sep,
	year = {2014},
	pages = {838--852},
}

@article{zechmeister_spectrum_2018,
	title = {Spectrum radial velocity analyser ({SERVAL}). {High}-precision radial velocities and two alternative spectral indicators},
	volume = {609},
	issn = {0004-6361},
	url = {http://adsabs.harvard.edu/abs/2018A%26A...609A..12Z},
	doi = {10.1051/0004-6361/201731483},
	urldate = {2020-08-15},
	journal = {Astronomy and Astrophysics},
	author = {Zechmeister, M. and Reiners, A. and Amado, P. J. and Azzaro, M. and Bauer, F. F. and Béjar, V. J. S. and Caballero, J. A. and Guenther, E. W. and Hagen, H.-J. and Jeffers, S. V. and Kaminski, A. and Kürster, M. and Launhardt, R. and Montes, D. and Morales, J. C. and Quirrenbach, A. and Reffert, S. and Ribas, I. and Seifert, W. and Tal-Or, L. and Wolthoff, V.},
	month = jan,
	year = {2018},
	pages = {A12},
}

@ARTICLE{stefansson_sub-neptune-sized_2020,
       author = {{Stef{\'a}nsson}, Gu{\dh}mundur and {Ca{\~n}as}, Caleb and {Wisniewski}, John and {Robertson}, Paul and {Mahadevan}, Suvrath and {Maney}, Marissa and {Kanodia}, Shubham and {Beard}, Corey and {Bender}, Chad F. and {Brunt}, Peter and {Clemens}, J. Christopher and {Cochran}, William and {Diddams}, Scott A. and {Endl}, Michael and {Ford}, Eric B. and {Fredrick}, Connor and {Halverson}, Samuel and {Hearty}, Fred and {Hebb}, Leslie and {Huehnerhoff}, Joseph and {Jennings}, Jeff and {Kaplan}, Kyle and {Levi}, Eric and {Lubar}, Emily and {Metcalf}, Andrew J. and {Monson}, Andrew and {Morris}, Brett and {Ninan}, Joe P. and {Nitroy}, Colin and {Ramsey}, Lawrence and {Roy}, Arpita and {Schwab}, Christian and {Sigurdsson}, Steinn and {Terrien}, Ryan and {Wright}, Jason T.},
        title = "{A Sub-Neptune-sized Planet Transiting the M2.5 Dwarf G 9-40: Validation with the Habitable-zone Planet Finder}",
      journal = {\aj},
         year = 2020,
        month = mar,
       volume = {159},
       number = {3},
          eid = {100},
        pages = {100},
          doi = {10.3847/1538-3881/ab5f15},
archivePrefix = {arXiv},
       eprint = {1912.00291},
 primaryClass = {astro-ph.EP},
       adsurl = {https://ui.adsabs.harvard.edu/abs/2020AJ....159..100S}
}

@INPROCEEDINGS{ninan_habitable-zone_2018,
       author = {{Ninan}, J.~P. and {Bender}, Chad F. and {Mahadevan}, Suvrath and {Ford}, Eric B. and {Monson}, Andrew J. and {Kaplan}, Kyle F. and {Terrien}, Ryan C. and {Roy}, Arpita and {Robertson}, Paul M. and {Kanodia}, Shubham and {Stefansson}, Gudmundur K.},
        title = "{The Habitable-Zone Planet Finder: improved flux image generation algorithms for H2RG up-the-ramp data}",
    booktitle = {High Energy, Optical, and Infrared Detectors for Astronomy VIII},
         year = 2018,
       editor = {{Holland}, Andrew D. and {Beletic}, James},
       series = {Society of Photo-Optical Instrumentation Engineers (SPIE) Conference Series},
       volume = {10709},
        month = aug,
          eid = {107092U},
        pages = {107092U},
          doi = {10.1117/12.2312787},
       adsurl = {https://ui.adsabs.harvard.edu/abs/2018SPIE10709E..2UN}
}

%%%%%%%%%%%%%%%%%%%%%%%%%%%%%%%%%%%%%%%%%%%%%%%%%%%%%%%
%%%%%%%%%%%%%%%%%%% Appendix %%%%%%%%%%%%%%%%%%%%%%%
%%%%%%%%%%%%%%%%%%%%%%%%%%%%%%%%%%%%%%%%%%%%%%%%%%%%%%%
\appendix
\section{Lack of activity correlation with HPF RVs}\label{app:activity}

\begin{figure}[!pht]
    \centering
    \includegraphics[width=0.9\columnwidth]{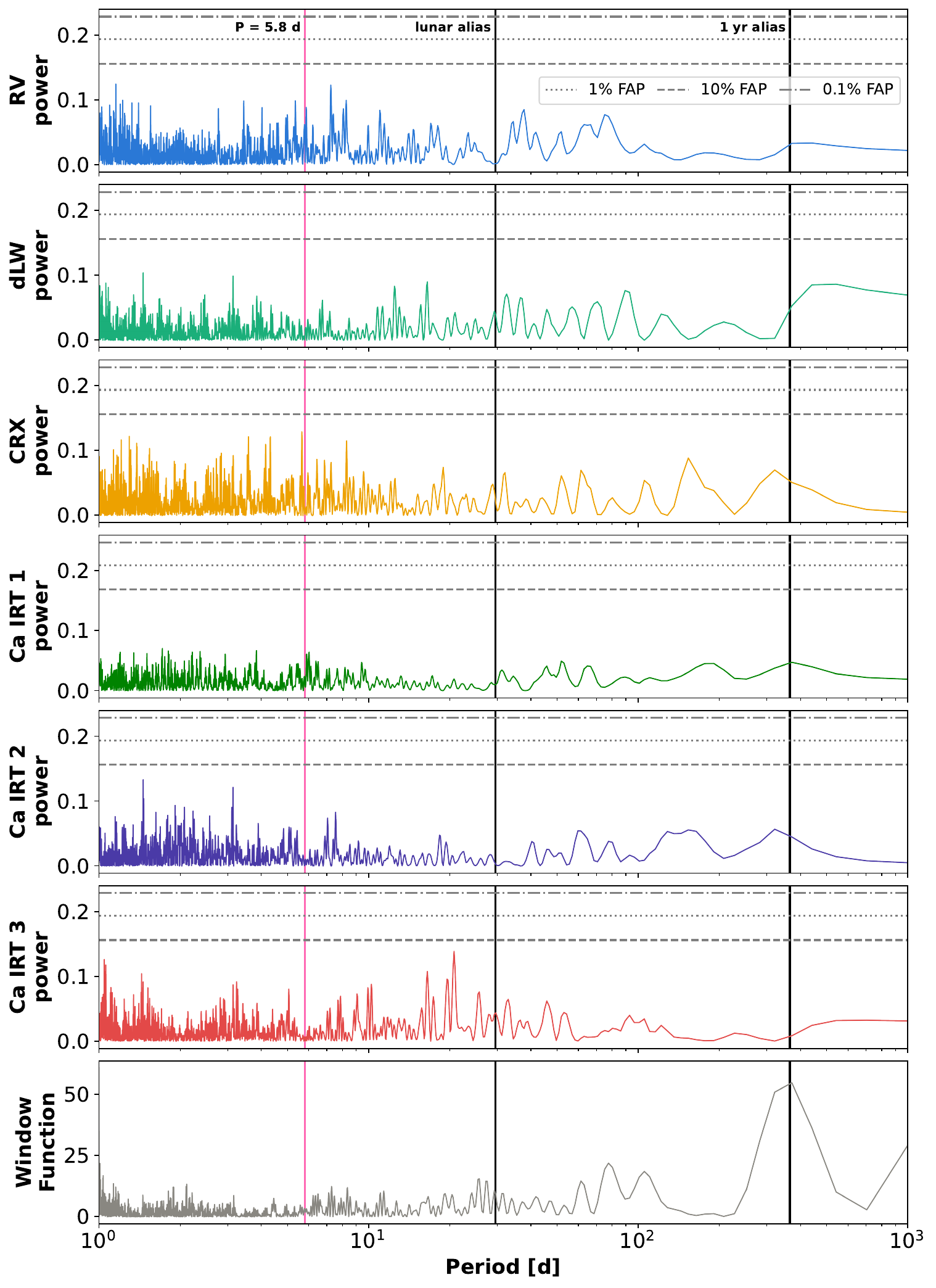}
    \vspace{-15pt}
    \caption{GLS periodograms of the HPF RVs, activity indicators, and window function, with the period of TOI-2120\,b indicated by the vertical orange line. The FAP levels corresponding to 0.1\%, 1\%, and 10\% are also marked.}
    \label{fig:activity_gls}
\end{figure}

\begin{figure}[ht]
    \centering
    \includegraphics[width=\columnwidth]{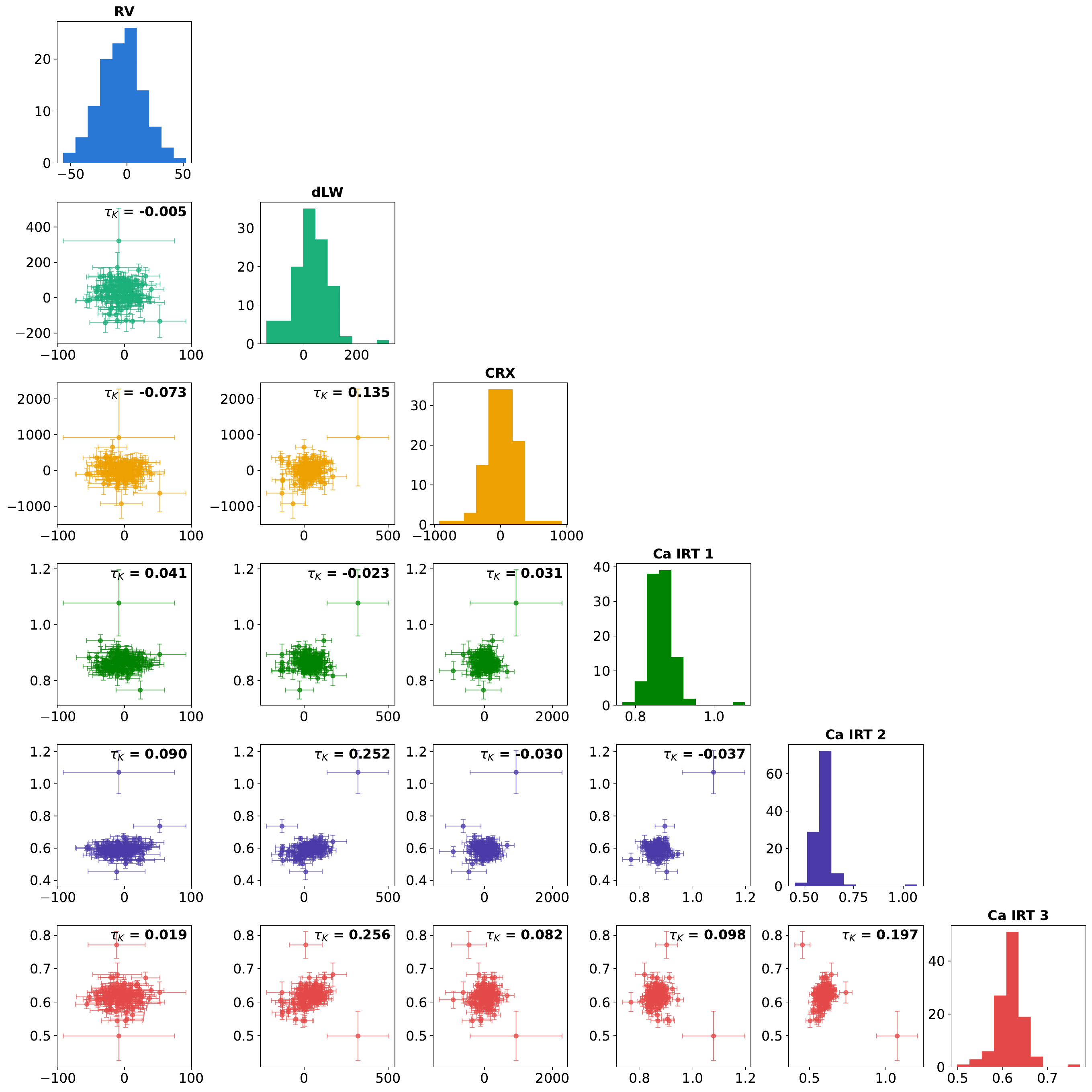}
    \caption{Corner plot of HPF RVs and activity indicators, with the Kendall $\tau$ coefficient included for each pair.}
    \label{fig:activity_corner}
\end{figure}

\clearpage
\section{Joint Fit of HPF and IRD}\label{app:ird}
We jointly model our HPF RVs and publicly available IRD measurements. The combined dataset consists of 38 HPF and 15 IRD observations. We fit the measurements simultaneously using a single Keplerian model and a linear background trend, allowing for independent instrumental zero-point offsets and jitter terms. The joint fit yields $K_b = 3.73^{+2.34}_{-2.07}$\,m\,s$^{-1}$, with inferred jitter amplitudes of 4.97\,m\,s$^{-1}$ for HPF and 12.18\,m\,s$^{-1}$ for IRD. The best-fit RV model and phase-folded orbital solution are presented in
Figure~\ref{fig:rvs_hpf_ird}.

\begin{figure}[!htp]
    \centering
    \includegraphics[width=\linewidth]{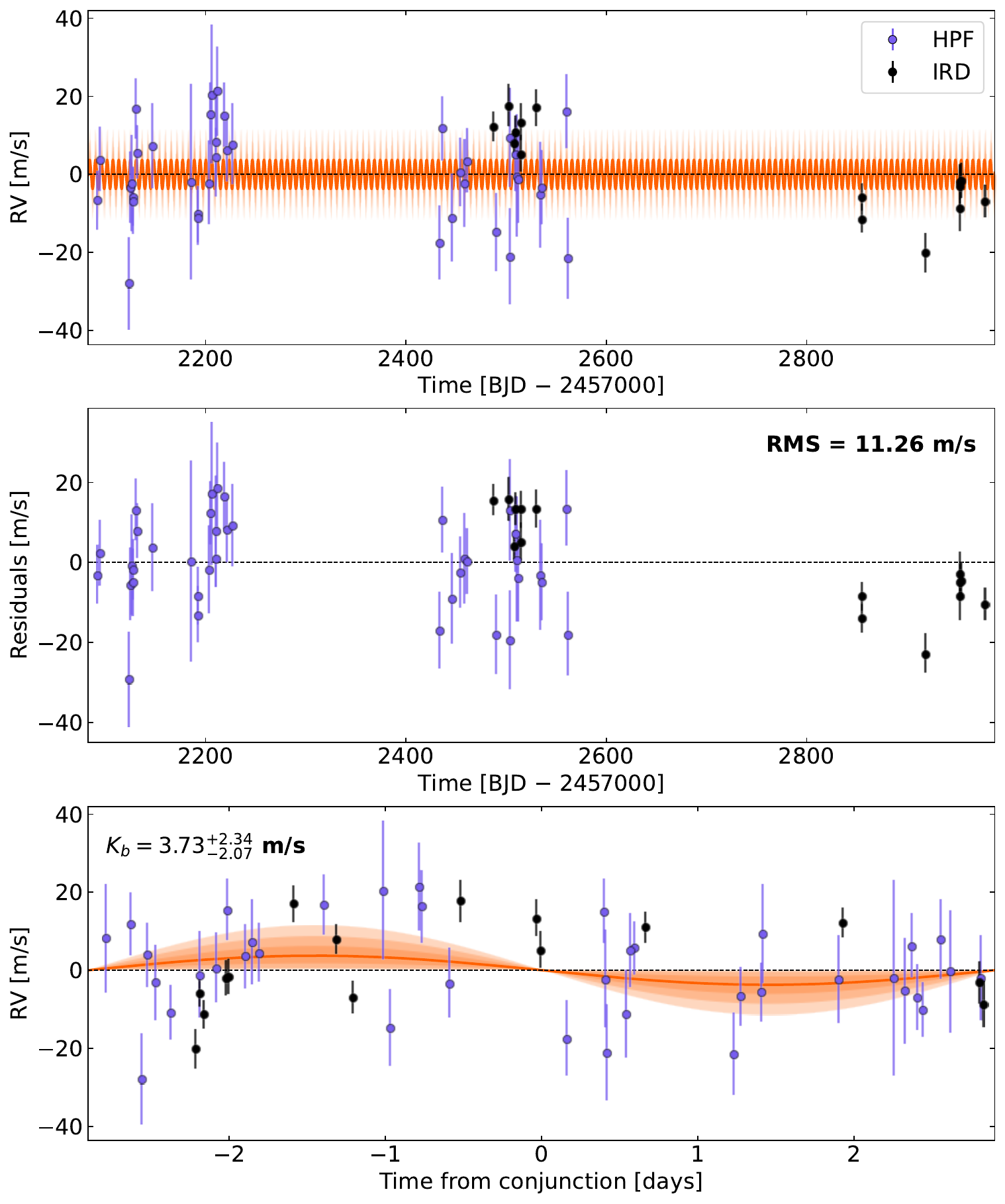}
    \caption{RVs of TOI-2120 measured with HPF (purple) and IRD (black), along with the joint best-fit model. \textit{Top:} RV time series with the per-instrument zero-point offsets removed. The orange line denotes the median model, comprising a Keplerian orbit and a linear trend, while the shaded orange regions indicate the 1, 2, and 3$\sigma$ credible intervals (darkest to lightest). \textit{Middle:} RV residuals relative to the median model, with an rms of 11.26\,m\,s$^{-1}$. \textit{Bottom:} Phase-folded RVs after removing the linear trend, with the Keplerian model and corresponding 1, 2, and 3$\sigma$ credible intervals. The inset reports the RV semi-amplitude, $K_b = 3.73^{+2.34}_{-2.07}$\,m\,s$^{-1}$. Error bars represent the 1$\sigma$ formal uncertainties}
    \label{fig:rvs_hpf_ird}
\end{figure}

\end{document}